\documentclass[fleqn,usenatbib]{mnras}

\usepackage[T1]{fontenc}
\usepackage{lineno}
\DeclareRobustCommand{\VAN}[3]{#2}
\let\VANthebibliography\thebibliography
\def\thebibliography{\DeclareRobustCommand{\VAN}[3]{##3}\VANthebibliography}

\usepackage{graphicx}	
\usepackage{amsmath}	
\usepackage{amssymb}	
\usepackage{lineno}
\usepackage{soul}
\usepackage{newtxtext,newtxmath}

\title[Star-clump encounters and exp. discs ]{Orbits and action changes during star-clump encounters
responsible for the origin of exponential discs in dwarf galaxies
 }
 
\author[J. Wu et al.]{
Jian Wu,$^{1}$\thanks{E-mail: jianwu@iastate.edu (JW), curt@iastate.edu (CS) }
Curtis Struck,$^{1}$\footnotemark[1]
Bruce G. Elmegreen$^{2}$
and Elena D'Onghia$^{3}$
\\
$^{1}$Department of Physics and Astronomy, Iowa State University, 2323 Osborn Dr., Ames, IA 50011, USA\\
$^{2}$IBM Research Division, T.J. Watson Research Centre, 1101 Kitchawan Road, Yorktown Heights, NY 10598, USA\\
$^{3}$Department of Astronomy, University of Wisconsin-Madison, 475 N Charter St, Madison, WI 53706, USA
}

\date{Accepted 2022 September 24. Received 2022 September 18; in original form 2022 July 20}

\pubyear{2022}

\begin{document}
\label{firstpage}
\pagerange{\pageref{firstpage}--\pageref{lastpage}}
\maketitle

\begin{abstract}
Previous studies found that stellar scattering by massive clumps can lead to the formation
of exponential profiles in galaxy discs, but details on how a star is moved around have not been fully explained. 
We use a {\small GADGET}-2 simulation where an exponential profile forms from an  initially Gaussian disc in about 4 Gyr for a low-mass galaxy like a dwarf irregular.  We find 
that nearly all large angular momentum changes of stars are caused by star-clump encounters with 
the closest approach less than 0.5 kpc. During  star-clump encounters, stars may increase their random motions, resulting in an increase in the average radial and vertical actions of the whole stellar population. 
The angular momentum change and the radial action change of an individual star are influenced by the direction from which 
the star approaches a clump. A star initially at a higher galactic radius relative to the scattering clump usually gets pulled inwards and loses its angular momentum during the encounter, and one at a lower radius
tends to shift outwards and gains angular momentum. The increase in the radial action
is the largest if a star encounters a clump from the azimuthal direction, and is the smallest from a
radial approach. The angular momentum change due to encounters  has an inward bias when the clump profile has a steep radial decline, and a shallow decline can make the bias outwards. The stellar profile evolution towards an exponential seems to occur regardless of the direction of the bias.

\end{abstract}




\section{Introduction}

Galaxies form by the collapse of cosmic filaments with frequent interactions
producing their individual angular momenta \citep{Nabb2017ARA&A..55...59N}. The non-dissipative component of
this collapse, e.g., dark matter, ends up in a thick halo and the dissipative
component, initially gas, forms a disc. Star formation in this gas builds up a
stellar disc over time, with high turbulent speeds at the earliest stages forming
a thick stellar disc \citep{Elmegreen2006ApJ...650..644E}, which is still faintly visible as old stars in galaxies
today \citep{Burstein1979ApJ...234..829B}, and gradually decreasing turbulence forming a thin disc that
dominates their visual appearance in optical images \citep{Yoachim2006AJ....131..226Y}.
During most of this process, the stellar disc has an exponential radial profile,
i.e., in both the thick and the thin components \citep{Comern2012ApJ...759...98C}, although there is often an
upward or downward break in this profile with an exponential on either side \citep{Pohlen2006,Bouquin2018ApJS..234...18B,Staudaher2019MNRAS.486.1995S}. The
fundamental origin of this exponential form is unclear. Explanations 
 range from
the initial collapse profile \citep{Mestel1963,Freeman1970ApJ...160..811F} to disc viscosity \citep{Lin1987} to spiral \citep{Sellwood2002} or bar \citep{hohl1971} torques, but all of these have problems \citep{Ferguson2001MNRAS.325..781F,Daniel2018MNRAS.476.1561D}, or are not universally applicable \citep{Elmegreen2013ApJ...775L..35E}, which is a key feature  of galaxy exponentials.  For example, models that rely on shear for viscosity or magnetic torques might not work in dwarf irregular galaxies, which have low shear yet extensive exponential stellar profiles \citep{Herrmann2013AJ....146..104H}.

Point scattering  of stars can make the radial
exponential profile \citep{Elmegreen2013ApJ...775L..35E,Elmegreen2014ApJ...796..110E,Elmegreen2016ApJ...830..115E,Struck2017MNRAS.464.1482S,Struck2017,Struck2019MNRAS.489.5919S}  at the same time as it increases the stellar velocity dispersion,
e.g., by cloud scattering, as in the proposal by \citet{Spitzer1951ApJ...114..385S}.
Studies of the dispersion alone \citep[e.g.,][]{Wielen1977} cannot predict the radial dependence of the scattered stars. To study the radial dependence without additional complications that arise in a real galaxy, other disc processes such as gravitational interactions between  individual stars and  spiral density waves  \citep{Shu2016ARA&A..54..667S}, were removed in several of our previous papers, leaving only
non-interacting  stars to scatter off co-orbiting mass
perturbations.  In these papers, the result always converged to an exponential stellar disc in about one Gyr.

While scattering alone makes physical and mathematical sense for the origin of galaxy exponentials, the ultimate test is a fully self-gravitating \textit{N}-body simulation of a galaxy with a live disc and halo, as that will include all of the dynamical processes that develop simultaneously, such as density waves.  In simulations using the  {\small GADGET}-2 code
\citep{2005gadget}, \citet{2020Wu} started with a non-exponential stellar profile and artificially added 200 point masses to drive the scattering. These point masses had a mass of $7.3\times10^6 M_{\sun}$ and  were identified with giant interstellar cloud complexes close to the maximum mass of Galactic molecular clouds \citep{Williams1997ApJ...476..166W}.
The initial vertical profile was assumed to be the equilibrium solution for an
isothermal layer, sech$^2(z/z_0)$, for vertical coordinate $z$ and scale height $z_0$. After slightly less than one billion years, the radial profile of the stellar disc always evolved to a
form of $e^{-r/h}/r$, which is indistinguishable from an exponential beyond one scale length.  The timescale was shorter for a less
radially stable disc, i.e., a lower Toomre parameter \citep{Toomre1964ApJ...139.1217T}, $Q$, and for more
massive clouds. At the same time as the exponential formed in surface density, the velocity
dispersion in the $z$-direction also developed an exponential radial profile with about twice the scale
length of the surface density. This factor of 2 results in a nearly constant disc
thickness, as also observed \citep{vander1981A&A....95..105V} but never explained  previously. In this model, disc evolution stops when the gas is gone.

All of these experiments including the full \textit{N}-body simulations 
gave exponential discs, but  questions  about how the orbital radius of each  star changes or why the changes lead to an exponential  were not examined. \citet{2020Wu} showed that a close encounter
between a star and a clump is able to modify the angular momentum of the star, but did not  indicate how much of the angular momentum change was caused by encounters.

Another interesting question is whether point scattering of stars exhibits a radial bias. If a bias exists, one may also ask what affects its direction and its strength. 
\citet{Elmegreen2016ApJ...830..115E} showed stochastic scattering in a disc with a constant inwards bias leads to an equilibrium near-exponential stellar profile. However, it is not clear that an inwards bias is required for the formation of an exponential if non-equilibrium states are acceptable.

In this paper, we try to fill the gaps described above and analyze star-clump encounters in  detail.
In Section~\ref{sec:f_model}, we introduce an \textit{N}-body simulation similar to the fiducial one of \citet{2020Wu} but with modifications
to further suppress unwanted perturbations in the initial disc. We also present
 a local shearing sheet model  to study orbits during encounter events.
In Section~\ref{sec:f_results_simulation}, we show results from the galaxy simulation, including the profile evolution
of the stellar disc, statistics of large stellar angular momentum changes, the connection between these changes and star-clump encounters, and action changes of stars during close encounters with clouds. Detailed explanations of these changes, based on properties of orbital action,  are provided in Section~\ref{sec:f_star_clump_enc}. Also in this section  we illustrate the orbits during star-clump encounters and use the orbits
to understand angular momentum changes and radial action changes of stars. We further show that the direction of scattering bias is related to the clump density distribution in the disc. A summary and conclusions are given in Section~\ref{sec:f_discuss_conclu}.

\section{Model}
\label{sec:f_model}

In this section, we explain properties of our \textit{N}-body simulations, methods used to calculate stellar actions, and the rotating frame in which star-clump encounters are studied by a set of differential equations for orbits in a shearing sheet.

\subsection{Simulations}
\label{sec:f_simula} 

\citet{2020Wu} showed, using a \textit{N}-body simulations, that galactic discs can evolve to an exponential due to star-clump interactions. To study the physical mechanisms responsible for the disc evolution, we set up a simulation using initial conditions similar to the fiducial run of \citet{2020Wu}, but with a few modifications.

In the simulation, the total mass of our galaxy is $1.45 \times 10^{11} M_{\sun}$, with 98\% of the total mass in the dark matter halo and 2\% in the stellar disc. The number of particles representing 
the dark matter and stars are $4 \times 10^{6} $ and $1 \times 10^{6} $
respectively.
The profile of the dark matter halo is a \citet{Hernquist1990} profile, and the density distribution  satisfies the following: 
\begin{equation}
    \rho_{\text{dm}}(r) =\frac{M_{\text{dm}}}{2\pi} \frac{a}{r(r+a)^3},
	\label{eq:dark matter dist}
\end{equation}
where $M_{\text{dm}}$ is the total mass of the halo and the Hernquist constant $a$ is set to 13.8 kpc. 
The initial density profile of the stellar disc is an isothermal distribution in the vertical direction and a Gaussian distribution in radius described by 
\begin{equation}
    \rho_{\text{star}}(r,z) =\frac{M_{\text{star}}}{2\pi z_0 h^2} ~ \text{sech}^2(\frac{z}{z_0 })~ e^{-r^2/h^2},
	\label{eq:star 3D dist}
\end{equation}
where  $M_{\text{star}}$ is the total mass of the stellar disc,
the scale length $h$ is set to 5.4 kpc, and the scale height $z_0$ is set to 0.45 kpc.  The scale length for the Gaussian corresponds to a smaller effective scale length for an exponential equivalent: if we consider a disc at 4 exponential scale lengths, where the surface density is $\exp(-4)$ times the central value, this position corresponds to a radius for the Gaussian where $(r/h)^2=4$, which means that the equivalent exponential scale length is $r/4=0.5h$, or 2.7 kpc. This is a fairly small galaxy, consistent with the stellar mass (see below) of $1.45\times10^9\;M_\odot$. 
The disc scale height of 0.45 kpc is consistent with the large ratios of thickness to length in dwarf irregular galaxies, where \cite{hunter06} found an average ratio of $\sim0.5$ for 94 dIrrs.
Initially, there are no clumps in the model. 

This model galaxy is evolved without clumps for 2.94 Gyr under the control of 
the \textit{N}-body package {\small GADGET}-2, designed by \citet{2005gadget}.  Without clumps, little evolution occurs in the profile of the stellar disc during this time. The purpose of the 2.94 Gyr is to let the disc settle down from the initial  conditions. A specific goal is to let ring waves 
in the stellar disc, due to initial imperfect centripetal balance, gradually
die out. 

At 2.94 Gyr, we reduce the mass of all the star particles by one half, and then  randomly select 210 of these particles to have their mass increased to $7.3 \times 10^{6} M_{\sun}$. We henceforth identify these particles as clumps \citep[similar to][]{D'Onghia2013ApJ...766...34D}. At this point, the total mass of all the clumps is roughly the same as the total mass of the stars, and the density profile of clumps is the same as that of stars as well. Like the stars, the clumps have a relatively large scale height, which is typical for the interstellar media of low-mass galaxies \citep{dalcanton04}. 

We run the galaxy with {\small GADGET}-2 for another 5.88 Gyr. In this period, stars interact with clumps, and the stellar profile changes significantly. When referring to time T elapsed in the simulation, we measure it from the beginning of this period. The  stellar profile at T = 0 is shown in Figure~\ref{fig:ff sdp change}. The Toomre Q profile at T = 0 and the component and total rotation curves at T = 0 are shown in Figures~\ref{fig:ff toomre Q} and ~\ref{fig:ff rot curve}, respectively.

A parameter comparison between this simulation and the fiducial run in \citet{2020Wu} is summarized  in Table~\ref{tab:para table}. A significant change is that dark matter mass is increased here in order to stabilize the stellar disc against  spontaneous spiral waves and wavelets induced by the clumps. Since these waves also scatter stars, the large halo mass suppresses their effect  on stellar orbits and makes star-clump interactions that we want to study more pronounced. Related to this change is an increase in the average value of the Toomre Q parameter (Fig.~\ref{fig:ff toomre Q}) compared to our previous work. The high value of Q (see Fig.~\ref{fig:ff toomre Q}) in the numerical experiments reduces the gravitational response of the disk to perturbations. 
Modest disk self-gravity 
also prevents wavelets from amplifying around the clumps \citep{Julian1966ApJ...146..810J,D'Onghia2013ApJ...766...34D}.
   N-body simulations of low-mass disks, perturbed by clumps and with lower $Q \approx 1.3$, 
were shown in \citet{D'Onghia2013ApJ...766...34D}. In those experiments, the disc started with an exponential shape so major structural changes could not be observed, the clumps were not allowed to move radially so they could not transfer their own angular momentum to stars, and the clumps were lower mass than in the present simulations. As a result, the structural parameters of the stellar disk did not change over time and there was no significant stellar heating as in the present simulations (see also \citet{Vera-Ciro2014ApJ}).

Other parameters related to stars and clumps  in this work are roughly the same as in \citet{2020Wu}, except the initial shape of clump density profile.  We will discuss the effect of this profile later in the paper. For parameters that are not explicitly listed in the table, we use the same values as \citet{2020Wu}. For example, gravitational softening lengths used in the model for dark matter, stars, and clumps are 0.2 kpc, 0.1 kpc, and 0.1 kpc, respectively. These values are identical to \citet{2020Wu}.

\begin{table*}
	\centering
	\caption{Parameter comparison between the model in this paper and the fiducial run in \citet{2020Wu}}
	\label{tab:para table}
	\begin{tabular}{lcc} 
		\hline
		Parameters & Values in this paper & Values in \citet{2020Wu}\\
		\hline
		Total mass $M_{\text{tot}}$ & $1.45 \times 10^{11} M_{\sun}$ & $2.9 \times 10^{10} M_{\sun}$\\
		Dark matter mass $M_{\text{dm}}$ & $1.42 \times 10^{11} M_{\sun}$ (98\% $M_{\text{tot}}$) & $2.6 \times 10^{10} M_{\sun}$  (90\% $M_{\text{tot}}$) \\
		Star mass $M_{\text{star}}$ & $1.45 \times 10^{9} M_{\sun}$ (1\% $M_{\text{tot}}$) & $1.45 \times 10^{9} M_{\sun}$  (5\% $M_{\text{tot}}$) \\
		Clump mass $M_{\text{clump}}$ & $1.52 \times 10^{9} M_{\sun}$ (1\% $M_{\text{tot}}$) & $1.45 \times 10^{9} M_{\sun}$  (5\% $M_{\text{tot}}$) \\
		
		Number of dark matter particles $N_{\text{dm}}$ & 4 000 000 & 970 000\\
		Number of star particles $N_{\text{star}}$ & 999 790 & 1 000 000\\
		Number of clumps $N_{\text{clump}}$ & 210 & 200\\
		
		Individual star mass $M_{\text{star}}$/$N_{\text{star}}$& $1.45 \times 10^{3} M_{\sun}$ & $1.45 \times 10^{3} M_{\sun}$  \\
		Individual clump mass $M_{\text{clump}}$/$N_{\text{clump}}$ & $7.3 \times 10^{6} M_{\sun}$ & $7.3 \times 10^{6} M_{\sun}$  \\
		Dark halo Hernquist constant $a$ & 13.8 kpc & 8.05 kpc\\ 
		Initial scalelength of the stellar disc $h$ & 5.4 kpc & 5.4 kpc\\
		Initial scale height of the stellar disc $z_0$ &  0.45 kpc & 0.45 kpc \\
		initial shape of clump density profile & $e^{-r^2/h^2}$ & $\frac{1}{\sqrt{r}}$ \\
		
		\hline
	\end{tabular}
\end{table*}

\subsection{Action calculation}
\label{sec:f_action_cal}

To investigate  the time evolution of stellar orbits, we study the angle-action variables. Actions are generally conserved integrals in fixed potentials, so their changes reflect perturbations in the stellar motions. The actions convenient to use in an axisymmetric potential are the radial action, vertical action, and azimuthal action. The specific radial action of a star is defined as
\begin{equation}
    J_r = \frac{1}{2\pi} \iint_V  \,dr\,dv_r ,
	\label{eq:raction}
\end{equation}
where $r$ and $v_r$  are the radial position and the radial velocity of the star, and $V$ is the enclosed region of the stellar orbit in  phase space. Similarly, the specific vertical action is defined as
\begin{equation}
    J_z = \frac{1}{2\pi} \iint_V  \,dz\,dv_z ,
	\label{eq:zaction}
\end{equation}
where $z$ and $v_z$ are the vertical position and the vertical velocity of the star. The specific azimuthal action is simply the $z$-component of the specific angular momentum of the star \cite[see Section 3.5 of ][]{2008book} , i.e.,
\begin{equation}
    J_\phi = L_z .
	\label{eq:phiaction}
\end{equation}

The {\small GADGET}-2 simulation generates positions and velocities of stars and clumps every 1.96 Myr. The actions of a star at a given moment are calculated based on its position, its velocity, and the galaxy potential at the moment, using the {\small AGAMA} software library developed by \citet{Vasiliev2019MNRAS.482.1525V}. 
Technically, the actions are path integrals around the whole of an orbit. The instantaneous actions from the {\small AGAMA} code  can be interpreted as path integrals
around a hypothetical orbit derived from the instantaneous position and velocity of a star.
The {\small AGAMA} library assumes that stellar orbits are integrable, and approximates the galaxy potential with a best fitting St{\"a}ckel potential when calculating the actions. This method is called `St{\"a}ckel fudge' \citep{Binney2012MNRAS.426.1324B}. The {\small AGAMA} function \verb'ActionFinder' is specially called to produce radial actions, vertical actions, and azimuthal actions.
We compared  the specific azimuthal actions obtained from \verb'ActionFinder' with the $z$-components of specific angular momenta $\mathbfit{L}$,
calculated by
\begin{equation}
     \mathbfit{L} = \mathbfit{r} \times \mathbfit{v},
	\label{eq:lz}
\end{equation}
where $\mathbfit{r}$ represents positions and $\mathbfit{v}$ represents velocities. The difference is typically less than 1 per cent. We will consider the results of these action calculations in Sec. 3.3.


\subsection{Motions of stars in a rotating frame: shearing sheet model}
\label{sec: f motion stars rotating frame}

In the {\small GADGET}-2 simulation, the ensemble of stars experience a slight change of orbital angular momentum on average when massive clouds are present. The reason for this loss is that individual stars change their angular momentum just because they interact with the clumps. Higher order effects may include or be related to a non-uniform clump density profile, orbital geometry in the local region of a star-clump encounter, scattering with a clump binary or ternary,  a velocity dispersion gradient in the stars, and other factors. To figure out the dominant factor, we also study star-clump encounters in a simplified model, assuming that the galaxy potential produces a flat rotation curve at all radii. We assume clumps rotate about the galactic centre in perfect circular orbits on the midplane of the galaxy and are not influenced by gravity from a star. 
When a star approaches a clump, the star feels a gravitational force from the clump and the force due to the galaxy potential gradient.  We assume the clump has a galactic radius $R_0$, and the constant circular speed of the galaxy is $v_0$.

To study the motion of the star, we choose a rotating frame centred on the clump with the $x$-$y$ plane coinciding with the midplane of the galaxy.  $\mathbfit{e}_x$ is along the radial direction of the galaxy, pointing away from the galactic centre. $\mathbfit{e}_y$ points towards the direction of galaxy rotation at the location of the clump. The angular velocity 
$\boldsymbol{\Omega}_0$ of this frame in the inertia{\color{green}l} frame is $\Omega_0~ \mathbfit{e}_z$, where the angular speed $\Omega_0$ satisfies
\begin{equation}
     \Omega_0 = \frac{v_0}{R_0}.
	\label{eq:omega}
\end{equation}
In the rotating frame, the coordinate of the galactic centre is $(-R_0, 0, 0)$. The equation of motion of the star is 
 \begin{equation}
    \frac{\text{d}^2 \mathbfit{x}}{\text{d}t^2}  = -\nabla\Phi - \nabla\Phi_{\text{clump}} - 2\boldsymbol{\Omega}_0 \times \frac{\text{d} \mathbfit{x}}{\text{d}t} - \boldsymbol{\Omega}_0 \times (\boldsymbol{\Omega}_0 \times (\mathbfit{x}+\mathbfit{p})),
	\label{eq:equ of motion}
\end{equation}
where $\mathbfit{x} = (x,y,z)$ is the coordinate of the star in the rotating frame, $\Phi$ is the galaxy potential,  $\Phi_{\text{clump}}$ is the gravitational potential of the clump, $\mathbfit{p} = (R_0, 0, 0)$ is a constant vector which makes the coordinate transition from the rotating frame centred on the clump to a rotating frame with the origin at the galactic centre.  Since the model galaxy has a constant circular speed, the galaxy potential gradient on the midplane satisfies
 \begin{equation}
    \nabla\Phi =  \frac{v_0^2}{|\mathbfit{x}+\mathbfit{p}|}
     \frac{\mathbfit{x}+\mathbfit{p}}{|\mathbfit{x}+\mathbfit{p}|},
	\label{eq:potential}
\end{equation}
which gives
\begin{equation}
     \nabla\Phi|_x =  \frac{v_0^2}{|\mathbfit{x}+\mathbfit{p}|}
     \frac{x+R_0}{|\mathbfit{x}+\mathbfit{p}|},
	\label{eq:potential x}
\end{equation}
\begin{equation}
     \nabla\Phi|_y =  \frac{v_0^2}{|\mathbfit{x}+\mathbfit{p}|}
     \frac{y}{|\mathbfit{x}+\mathbfit{p}|}.
	\label{eq:potential y}
\end{equation}
The clump potential gradient satisfies
\begin{equation}
    \nabla\Phi_{\text{clump}} =  \frac{G m_{\text{clump}}}{|\mathbfit{x}|^2}
     \frac{\mathbfit{x}}{|\mathbfit{x}|},
	\label{eq:potentia clumpl}
\end{equation}
where $m_{\text{clump}}$ is the mass of individual clump, and $G$ is the gravitational constant.
 We constrain the motion of the star to the midplane of the galaxy, so the $z$ component of galactic gravity is ignored. Equation~\ref{eq:equ of motion} then reduces to  two-dimensions and becomes the following:
 \begin{equation}
    \frac{\text{d}^2 x}{\text{d}t^2}  = -\frac{v_0^2(x+R_0)}{(x+R_0)^2+y^2} - \frac{G m_{\text{clump}} x }{(x^2+y^2)^{3/2}}
    + 2\Omega_0 \frac{\text{d}y}{\text{d}t} 
   +\Omega_0^2(x+R_0),
	\label{eq:equ of motion x}
\end{equation}
 \begin{equation}
    \frac{\text{d}^2 y}{\text{d}t^2}  = -\frac{v_0^2 y}{(x+R_0)^2+y^2} - \frac{G m_{\text{clump}} y }{(x^2+y^2)^{3/2}}
    - 2\Omega_0 \frac{\text{d}x}{\text{d}t} 
   +\Omega_0^2 y.
	\label{eq:equ of motion y}
\end{equation}
These are the differential equations for the the local shearing sheet model.

In  Equations~\ref{eq:equ of motion x} and \ref{eq:equ of motion y}, $v_0$ is set to 113 km\,s$^{-1}$, which is the rotation speed of the outer disc in the {\small GADGET}-2
simulation; $m_{\text{clump}}$ is $7.3 \times 10^{6} M_{\sun}$, which is the mass of an individual clump. 
A gravitational softening length $s$ of 0.1 kpc is added to the 
denominator of the second term on the right-hand-side of Equations~\ref{eq:equ of motion x} and \ref{eq:equ of motion y}, so the denominator becomes ${(x^2+y^2+s^2)^{3/2}}$ instead of ${(x^2+y^2)^{3/2}}$. This softening length is the same as the softening length used in  {\small GADGET}-2; the purpose here is to model the extended size of a cloud. To obtain the trajectory of a star, Equations~\ref{eq:equ of motion x} and \ref{eq:equ of motion y} are solved numerically, using an explicit Runge-Kutta (4,5) formula \citep{Dormand1980, Shampine1997}.  Specifically, the \verb'ode45' function provided by {\small  MATLAB} is employed.

\section{Results from the galaxy simulation}
\label{sec:f_results_simulation}

\subsection{Time evolution of the simulated galaxy disc}
\label{sec:f_time_evo_simula}

Before presenting  the results from our analysis of orbits, we show in this subsection the macroscopic changes of the simulated disc with time, and compare it with the fiducial run in \citet{2020Wu}.

\begin{figure}
	\includegraphics[width=\columnwidth]{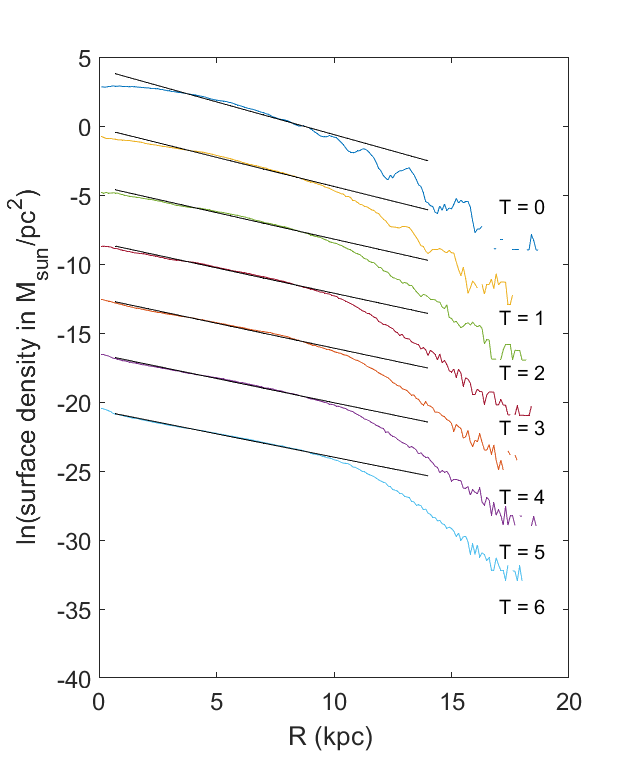}
    \caption{Evolution of the stellar surface density profile in the simulated galaxy. 
Except for T = 0, curves are offset downwards successively by 4 units to give clear views.  Black lines are the lines of best fit in the interval of  [2, 10] kpc. The R-squared of the linear fitting from T = 0 to T = 6 are 0.9759, 0.9835, 0.9849, 0.9906, 0.9873, 0.9929, and 0.9943, respectively.
The root mean squared error of the fitting from  T = 0 to T = 6 in units of the vertical axis are 0.174, 0.128, 0.111, 0.084, 0.096, 0.069, and 0.060, respectively. When using the slope of the best fit line to determine an exponential scale
  length, the effective scale length from T = 0 to T = 6 are 2.11, 2.36, 2.60, 2.73, 2.77, 2.85, and 2.96 kpc, respectively.  The time unit used in this plot is 0.98 Gyr.
    }
    \label{fig:ff sdp change}
\end{figure}

\begin{figure}
	\includegraphics[width=\columnwidth]{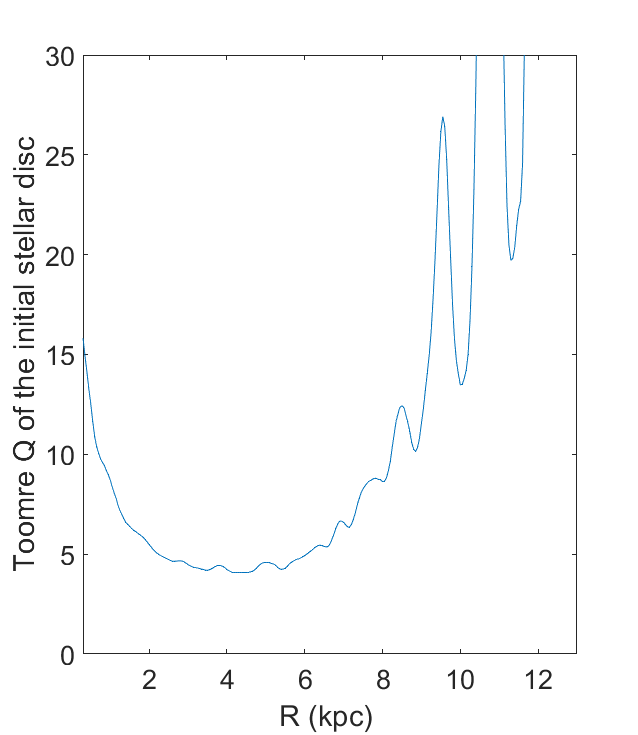}
    \caption{ Toomre Q of the stellar disc at T = 0. The oscillation in the curve beyond $R = 8$ kpc is due to ring waves in stellar density.
    }
    \label{fig:ff toomre Q}
\end{figure}

\begin{figure}
	\includegraphics[width=\columnwidth]{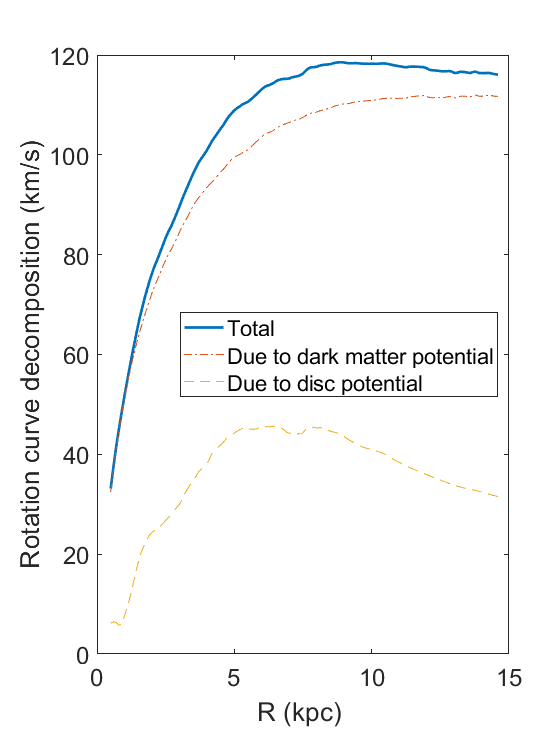}
    \caption{ Azimuthally averaged rotation curve of the disc at T = 0. The solid blue curve shows the rotational speeds on the galaxy midplane. The dash-dotted orange curve and the dashed yellow curve give contribution due to dark matter potential gradient and disc potential gradient, respectively.   
    }
    \label{fig:ff rot curve}
\end{figure}

Figure~\ref{fig:ff sdp change} shows the time evolution of the surface density profile of the stellar disc. At T = 0,   
the stellar disc has a Gaussian profile with a few ring waves outside of $R = 10$ kpc. Later on, these ring waves die out, and the profile within $R = 10$ kpc evolves to a nearly straight line, which represents an exponential disc. At T = 6, the whole disc is a Type II exponential with a small overdense region at the galactic centre.

This profile evolution confirms the formation of an exponential disc out of star-clump scattering as seen in \citet{2020Wu}. Since the initial clump distribution used here is different from that of \citet{2020Wu}, this evolution result also indicates that 
the profile change towards an exponential through star-clump scattering happens with various clump density distributions. However,  the clump distribution can affect the scalelength of the exponential formed. When comparing the final stellar profile with that of \citet{2020Wu}, a difference is that the density profile has 
a steeper slope or a smaller scalelength than \citet{2020Wu} in the region outside of $R = 10$ kpc. This can be 
explained by the clump density distinction. As shown in Table\ref{tab:para table}, \citet{2020Wu} used  a $1/\sqrt{r}$ clump radial profile, while the run in this paper has an $e^{-r^2/h^2}$ profile. 
The number of clumps in the latter decreases  with increasing radius faster than the former.
Therefore, fewer clumps are outside of $R = 10$ kpc, compared with the former run. Owing to limited star-clump interaction, the stellar profile in this paper  has little change outside of $R = 10$ kpc despite the disappearance of ring waves.

 We also recall that a model with a flat initial density profile was described in \citet{2020Wu}. Other initial profiles were tested in unpublished models. Qualitatively, the profile evolution towards an exponential form is not changed by these different initial conditions.

\begin{figure}
	\includegraphics[width=\columnwidth]{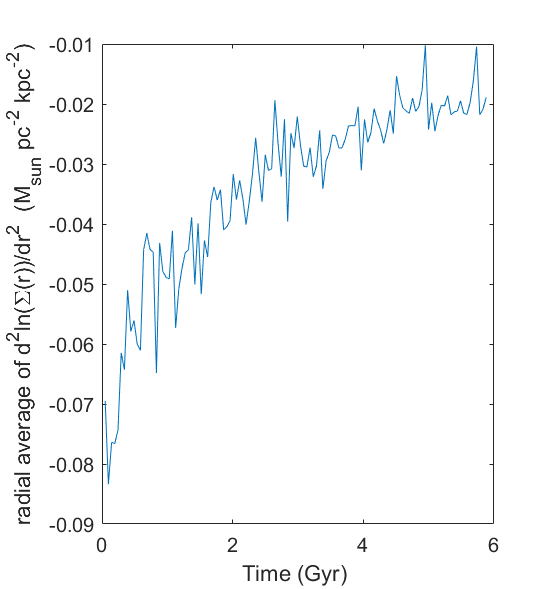}
    \caption{Time evolution of the second derivative of the logarithm radial stellar profile. 
The radial average is taken from 2 kpc to 9 kpc.
    }
    \label{fig:ff sec deriv}
\end{figure}

To better quantify how closely a profile fits an exponential, we calculate the second derivative of the logarithmic 
density profile with respect to the radius $R$, and then take the average value from $R = 2$ to $R = 9$ kpc.
When computing the second derivative, we firstly divide the disc into concentric annuli, each with a width of 1 kpc. Secondly, the number of stars in every annulus is counted and then used to calculate the surface density of the annulus. Thirdly, the logarithm of the surface density is calculated and the second derivative at the radius of the annulus is approximated by   $(\text{ln} \Sigma (r + \delta r) + \text{ln} \Sigma (r - \delta r) - 2 \text{ln} \Sigma (r - \delta r) )/\delta r^2$, where the step $\delta r$ is 1 kpc. From $R = 2$ to $R = 9$ kpc, the values of the second derivative may fluctuate, but usually have the same sign.  
For a perfect exponential profile, the logarithm  of the radial density  equals $-r/h$, and the second derivative  of this is 0.
For a Gaussian profile with $e^{-r^2/h^2}$, the logarithm of the radial density  is $-r^2/h^2$, and the 
second derivative is $-2/h^2$. Here, the initial stellar disc is a Gaussian with $h = 5.4$ kpc, so the theoretical second
derivative is -0.069 $M_{\sun}$ pc$^{-2}$ kpc$^{-2}$.
Figure~\ref{fig:ff sec deriv} shows the time evolution of the second derivative for the simulated stellar disc. The initial value is around -0.07 $M_{\sun}$ pc$^{-2}$ kpc$^{-2}$, which agrees with the theoretical calculation. After a transitory drop,
the second derivative increases steadily and reaches -0.02 $M_{\sun}$ pc$^{-2}$ kpc$^{-2}$ at 5.88 Gyr.
The change of the second derivative towards 0 verifies that the disc is gradually turning  into an exponential profile.  The effective exponential scale length at 5.88 Gyr is 2.96 kpc, and smaller at earlier times.
The plot also shows that the rate of change is fast during the first 2 Gyrs, and slows down afterwards. 
When we analyze stellar orbits later in this paper, we will focus on dynamics in the first 2 Gyrs.

\begin{figure}
	\includegraphics[width=\columnwidth]{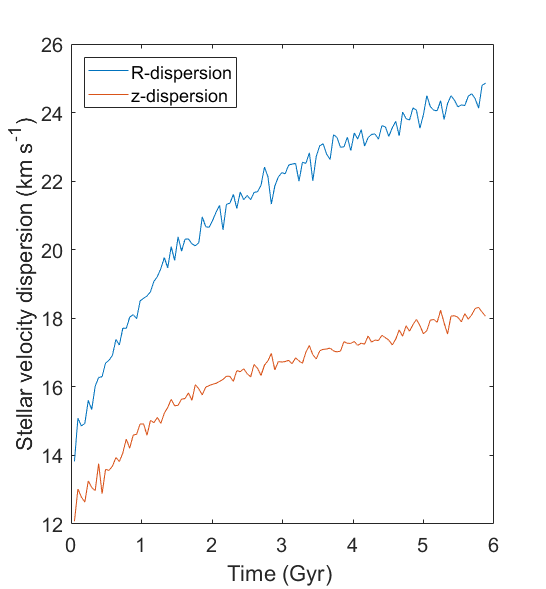}
    \caption{Time evolution of stellar velocity dispersions in the $r$-direction and $z$-direction. 
The velocity dispersions at each time point are valued at the galactic radius of 5 kpc.
    }
    \label{fig:ff vdisp}
\end{figure}

Figure~\ref{fig:ff vdisp} shows the time evolution of  the stellar velocity dispersions in the $r$-direction and $z$-direction  measured at a galactic radius of 5 kpc (i.e., over a range of 4.75-5.25 kpc). The $r$-dispersion goes from 14 km\,s$^{-1}$ to 25 km\,s$^{-1}$, and the $z$-dispersion goes from 12 km\,s$^{-1}$ to 18 km\,s$^{-1}$. The shape of the two curves resembles the fiducial run 
in \citet{2020Wu}. As we will see later in this paper, the increased random motion in the
$r$-direction and the $z$-dispersion can be explained by star-clump encounters.

\subsection{Stellar angular momentum change in the simulated galaxy}
\label{sec:f_stell_ang}

The change of the stellar density profile seen in Figure~\ref{fig:ff sdp change} is a collective effect of numerous changes in the orbital radius of individual stars. To
understand the main causes of the stellar profile change, we study angular
momentum changes of individual stars.

Changes in angular momentum of individual stars are evaluated in 
time windows with a duration of 196 Myr. A series of time windows, successively shifted by 19.6 Myr, is observed . Within a
time window, the change in specific angular momentum is defined as the 
difference between the maximum and the minimum angular momenta, and the time  for this angular momentum change is defined as the average of the two time points when
the maximum and the minimum occur.
We say a change in specific angular momentum is large when the value of the change is greater than   
75 kpc\,km\,s$^{-1}$. Given the rotation velocity of 115 km\,s$^{-1}$ in
the outer part of the stellar disc, this large specific angular momentum change indicates a radial shift of more than 0.65 kpc for a star.

\begin{figure}
	\includegraphics[width=\columnwidth]{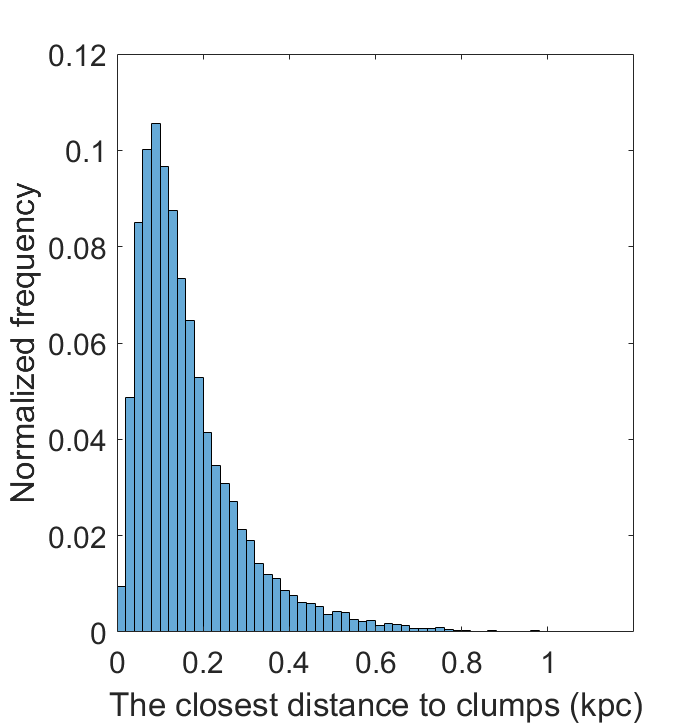}
    \caption{Distribution of the closest distance to any clumps during a 196 Myr time window when 
    a star experiences a $L_z$ change greater than 75 kpc\,km\,s$^{-1}$.
    The width of a bin in the plot is 0.02 kpc.
    }
    \label{fig:ff distance distri}
\end{figure}

In the simulated galaxy, the angular momenta of 9903 random stars are tracked. 
Figure~\ref{fig:ff distance distri} shows the distribution of the closest distance of a tracked star to any of the
clumps in a time window during which the tracked star has a large angular 
momentum change. 
The normalized frequency peaks at 0.1 kpc.
When adding up the bins where the closest distance is less than 0.5 kpc, a
probability of 0.973 is obtained. This indicates that 97.3\% of the large  stellar
angular momentum changes are associated with clump encounters with a 
distance less than 0.5 kpc. Therefore, close star-clump encounters are the main 
cause of stellar  angular momentum changes in the simulated galaxy.

In the first 1 Gyr of disc evolution, large angular momentum changes occur 11974 times for the 9903 stars. On average, each star experiences a large change 1.2 times in the first 1 Gyr.
The stars lose angular momentum in 49.23\% of the cases and gain in 50.77\%, so moving radially outwards
after a clump encounter is slightly favoured. However, this  result is not
statistically supported at a significance level of 0.05. 
Take a null hypothesis to be that the probability of gaining angular momentum is 0.5, and the
alternative hypothesis  to be that the probability is not equal to 0.5.
Considering the sample size and assuming the mean probability follows a normal distribution by applying the central limit theorem, the $p$-value of the
hypothesis testing is 0.094.
Because the $p$-value is greater than 0.05, the null hypothesis is not rejected.  This means that the probability of gaining or losing angular momentum could be 50\%, without a bias in one direction or the other. The topic on the radial bias of angular momentum changes will be discussed in detail later in this paper.

\begin{figure}
	\includegraphics[width=\columnwidth]{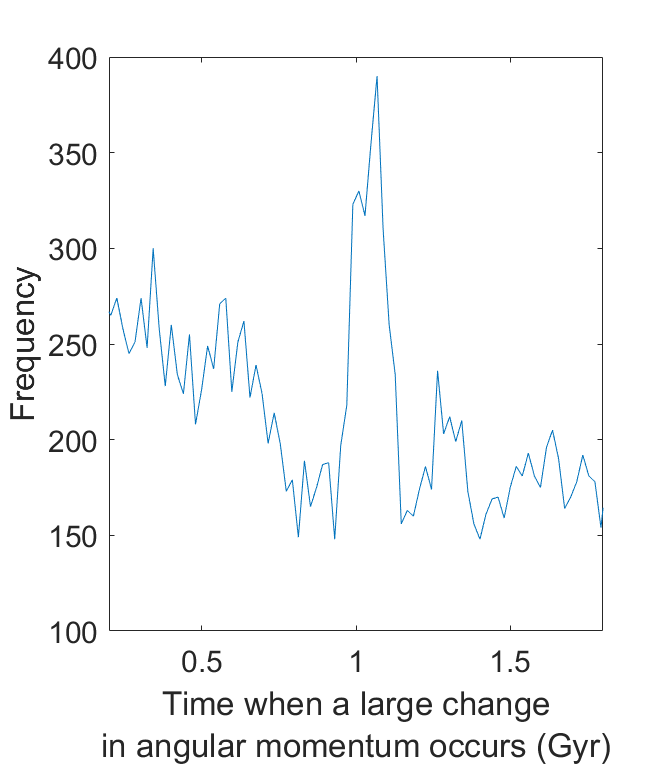}
    \caption{
    The frequency of large $L_z$ changes for 9903 tracked stars as a function of time from 0.2 Gyr
    to 1.8 Gyr. The bin width in the horizontal direction is 19.6 Myr.
    }
    \label{fig:ff time distri}
\end{figure}

Figure~\ref{fig:ff time distri} shows the frequency of large angular momentum changes in the radial interval from 2 to 9 kpc as a function of time from 0.2 Gyr to 1.8 Gyr. The bin width of the figure is 19.6 Myr. The plot has a large fluctuation near 1 Gyr, but exhibits an overall downward trend. (At this time the clumps are well distributed throughout the stellar disc, and thus, close to a maximum number of stars. At other times, the clumps are more clumped, and thus, close to fewer stars. Although large, this is essentially a statistical fluctuation.) The downward trend qualitatively matches the rate of stellar profile changes shown
on Figure~\ref{fig:ff sec deriv}. When the frequency of large $L_z$ changes decreases with time, the profile evolution slows down.

\begin{table*}
	\centering
	\caption{Mean action changes of  a star during one encounter with a clump at different Galactic radii in the  {\small GADGET}-2 run. An action change is calculated using the action value when a star enters a sphere of 0.5 kpc radius around the clump and the one when the star goes out of the sphere. 
	The galactic radius is the distance of the star to the galactic centre, evaluated at the moment when the star goes into the sphere. 
}
	\label{tab:Lch table}
	\begin{tabular}{ccccc} 
		\hline
		Galactic radius & Mean $L_z$ change  & Mean $J_r$ change 
		 & Mean $J_z$ change & Average number of encounters experienced  \\
(kpc)&	(kpc\,km\,s$^{-1}$) &(kpc\,km\,s$^{-1}$) &(kpc\,km\,s$^{-1}$) & in one gigayear per star\\
		\hline
			1 & -0.18 $\pm$ 0.03 & 0.09 $\pm$ 0.01  & 0.07 $\pm$ 0.01 & 77.6\\
	2 & -0.49 $\pm$ 0.04 & 0.06 $\pm$ 0.01  & 0.11 $\pm$ 0.01 &
	51.4\\
		3 & -0.51 $\pm$ 0.06 & 0.13 $\pm$ 0.01  & 0.11 $\pm$ 0.01
		& 35.2\\
		4 & -0.41 $\pm$ 0.09 & 0.23 $\pm$ 0.01  & 0.13 $\pm$ 0.01
		& 26.6\\
	5 & -0.62 $\pm$ 0.14 & 0.30 $\pm$ 0.02  & 0.19 $\pm$ 0.01
	& 16.6\\
		6 & -0.70 $\pm$ 0.25 & 0.34 $\pm$ 0.03  & 0.26 $\pm$ 0.02
		& 9.4\\
			7 & -0.45 $\pm$ 0.42 & 0.41 $\pm$ 0.05 & 0.34 $\pm$ 0.03 & 6.2 \\
				8 & -0.48 $\pm$ 0.72 & 0.64 $\pm$ 0.07 & 0.55 $\pm$ 0.05 & 3.4\\
				9 &  -0.64 $\pm$ 1.35 & 0.88 $\pm$ 0.14 & 0.69 $\pm$ 0.08 & 1.6\\
					10 &  -4.65 $\pm$ 3.20 & 1.05 $\pm$ 0.28 & 0.90 $\pm$ 0.15  & 0.6\\
		\hline
	\end{tabular}
\end{table*}

\subsection{Action changes during star-clump encounters}
\label{sec:f_action_change}

In the previous subsection, we showed that  the closest star-clump encounters are responsible for nearly all of the large $L_z$ changes in the simulation. This is true even with the softening parameter equal to 0.1 kpc for the cloud interactions. In this subsection,
we focus on the influence of these encounters on stellar action integrals.

In the simulated galaxy, trajectories of 9903 random stars are tracked with 
a high time resolution. 
We say a close encounter event occurs if any of the tracked stars enters 
the neighbourhood of any of the 210 clumps. Here, we define the neighbourhood
of a clump as the three-dimensional sphere centred on the clump with a radius
of 0.5 kpc. In the first 1 Gyr, 276 711 encounter events are detected for all the tracked stars. On average, each star experiences encounter events 27.9 times.
This is much more than the frequency of large $L_z$ changes shown in the
previous subsection. A large $L_z$ change is almost certainly caused by 
a close encounter, but an encounter may not produce a large $L_z$ change.
Also note that a star can reside in the neighbourhood of more than one
clump simultaneously. In this scenario, we count the event multiple times based
on the number of clumps involved. Encounters with multiple clumps usually happen  in the central region of the disc, since the clump density there is high.

Table~\ref{tab:Lch table} shows mean stellar action changes during one star-clump encounter  at
different radial locations in the first 1 Gyr. The action changes are calculated 
using the values when the star enters and leaves a clump neighbourhood. The
galactic radii in the table are radii of stars, evaluated when the stars go 
into the clump neighbourhood.
The last column in the table shows the average number of encounters experienced in 1 Gyr per star. As the clump surface density peaks at the galactic center, the average number of encounters per star increases as the galactic radius goes down. Although the mean action changes in one star-clump interaction have a small magnitude, frequent encounters can lead to a big effect.

From 1 kpc to 6 kpc, the mean $L_z$ change is negative, indicating 
that encounters are more likely to make a star lose $L_z$ and
move inwards after meeting with  a clump. This movement leads to an increase in stellar
density in the disc centre, and eventually drives the formation of
a central cusp as seen in Figure~\ref{fig:ff sdp change} at T = 6.
From 7 kpc to 10 kpc, the uncertainty in $L_z$ change is comparable to 
the absolute value. Therefore, the radial bias is not statistically significant.
This agrees with the null hypothesis in the previous subsection.
Large $L_z$ changes studied before are ones with changes of more than
75 kpc\,km\,s$^{-1}$. (Note: the $L_z$ changes in Table 2 are averaged over all stars, not just those with large changes.) They are unlikely to occur in the inner disc, because 
the rotation velocity there is low and a star has to radially migrate by 
a long distance to meet the criterion of 75 kpc\,km\,s$^{-1}$.
 The previous statistics on large $L_z$ changes mostly reflect scattering in the outer disc.

As shown in the table, the mean $J_r$ and $J_z$ changes are all positive, and the
values of changes generally increase with galactic radii. This means that encounters with clumps, on average, intensify the random motion of stars
in the $r$-direction and $z$-direction. It also explains the rises in the velocity
dispersion seen in Figure~\ref{fig:ff vdisp}.

\begin{figure*}
	\includegraphics[width=\textwidth]{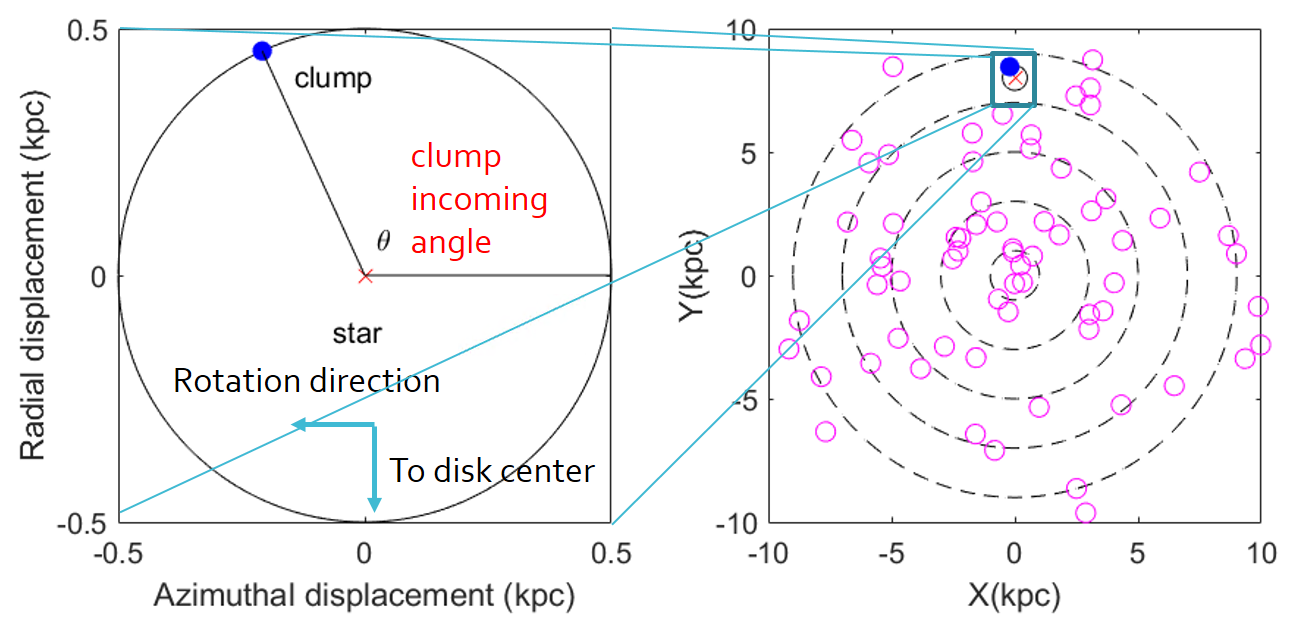}
    \caption{A demonstration of the definition of a clump incoming angle.
    Assume a star, represented by a red cross, enters the neighbourhood of a clump
    represented by a blue dot.
    The left panel shows a local reference frame centred on the star where the galactic centre is at the downward direction and the direction of disc rotation is the left. The clump incoming angle
    is the angular displacement of the clump in this frame, shown as the angle $\theta$ on the plot. The black circle has a radius of 0.5 kpc. If the clump has the same $z$-coordinate as the star when the star enters its neighbour, the clump will be on
    the circle as shown here. Otherwise, it will be inside the circle. If the star
    is at the galactic radius of 8 kpc, right panel shows the positions of the star and the clump along with the black circle in the global frame taking the galactic centre as the origin. Magenta dots in the plot represents other clumps
    in the galaxy. Concentric circles, added for clarity, are at 1 kpc, 3 kpc,
    5 kpc, 7 kpc, and 9 kpc. The galaxy rotates counterclockwise on this panel. The clump
    population shown on the plot is for demonstration purpose only, and does not reflect the clump density profile in the simulation.
    }
    \label{fig:ff demo angle}
\end{figure*}

To better understand the action changes during encounters, we classify encounter events based on clump incoming angles. The definition of a clump incoming angle is shown in Figure~\ref{fig:ff demo angle}. The left panel of the figure shows a face-on view of a local region in the galactic disc.  The red cross and the blue dot represents the locations of a star
and a clump, respectively. This local frame is centred on the star, downwards is the direction to the galactic centre, and left is the direction of disc rotation. This frame, with the heavy clumps orbiting past the light, but fixed star, is somewhat counter-intuitive initially, but it is the natural frame for the present calculations. We will use a clump-centred frame in other discussions below. 
When a star enters the neighbourhood of a clump, the clump will be on or inside the 0.5 kpc circle shown in the figure. The clump incoming angle $\theta$ is defined as the angle
measured counterclockwise from the right as shown in the figure. Specifically, if the clump approaches the star from below (negative y-value in the figure), then the angle span is measured clockwise and we define the clump incoming angle as a negative value. That is, we consider an angular range of from -180$^\circ$ to +180$^\circ$. Thus in general, if a clump
incoming angle is positive, the clump comes with a galactic radius greater
than the star, but if the angle is negative, the clump comes with a galactic
radius less than the star. The right panel of the figure shows the position of star-clump encounter region on the disc plane if the galactic radius of the star is 8 kpc. In this panel, the origin is the disc centre and the disc rotates counterclockwise. Magenta dots represent other clumps in the galaxy.

\begin{figure*}
	\includegraphics[width=\textwidth]{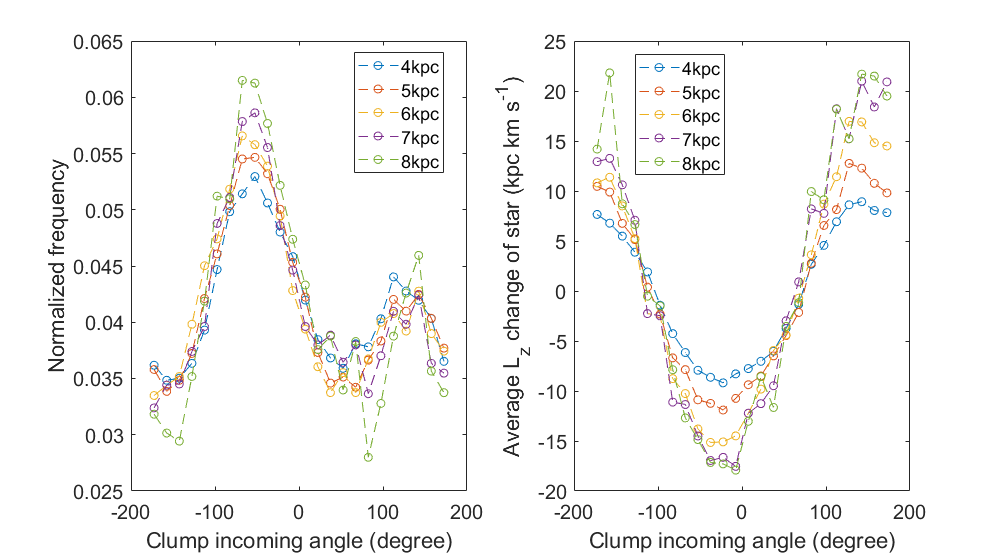}
    \caption{The distribution of clump incoming angles and average $L_z$
    changes as a function of clump incoming angles in star-clump collisions. Colors reflect
    encounters occurring at different galactic radii. The bin width in both panels is 15 degrees.
    }
    \label{fig:ff L_z and angle}
\end{figure*}

In Figure~\ref{fig:ff L_z and angle}, the left panel shows the distribution of clump incoming angles for encounter events occurring at different galactic radii. The 
right panel shows the average $L_z$ change of the encounters that 
belong to the same incoming angle bin. The width of the angle bins is 15 degrees.
We show encounters from 4 kpc to 8 kpc (over the first 2.0 Gyr), because this region has a sufficiently large encounter sample and does
not greatly suffer from encounters involving multiple clumps.

Curves on the left panel overlap and give two crests and two troughs.
This can be explained by the differential rotation of the disc.
Assume a star and a clump rotate about the galactic centre in circular
orbits. If the clump has a greater orbital radius than the star, it 
will have a smaller angular velocity. The encounter with the star
will occur when the star catches up with it, and this will lead 
to a clump incoming angle between 90$^\circ$ and 180$^\circ$.
If the clump has a smaller orbital radius than the star, it will
have a greater angular velocity. The encounter will occur when
it overtakes the star, and this gives a clump incoming angle between -90$^\circ$ and 0$^\circ$.
The angle intervals [-180$^\circ$, -90$^\circ$] and [0$^\circ$, 90$^\circ$]
are forbidden under the assumption of circular orbits.
When the star and the clump have some random energy in the radial direction, the two forbidden intervals can get filled, and this
 occurs in the two troughs on the plot.

Curves on the right panel also manifest a clear shape. 
Encounters with the clump incoming angle between -105$^\circ$ and
75$^\circ$ lose $L_z$ on average, and the rest of the angles gain. 
When the star and the clump have low orbital energy in random motion,
the trajectory of the star in the neighbourhood of the clump can be inferred
given a clump incoming angle, and the $L_z$ change of the
star can be determined from the trajectory. As a result, there is a
direct connection between the clump incoming angle and the $L_z$ change  for near-circular orbits. Clumps that come in between -105$^\circ$ and
75$^\circ$ initially have a lower orbital radius than the star and end up with a higher orbital radius, passing behind the star in azimuth. This pulls the star back in its orbit and causing it to lose angular momentum.
We will discuss this more in the next section. 

\begin{figure*}
	\includegraphics[width=\textwidth]{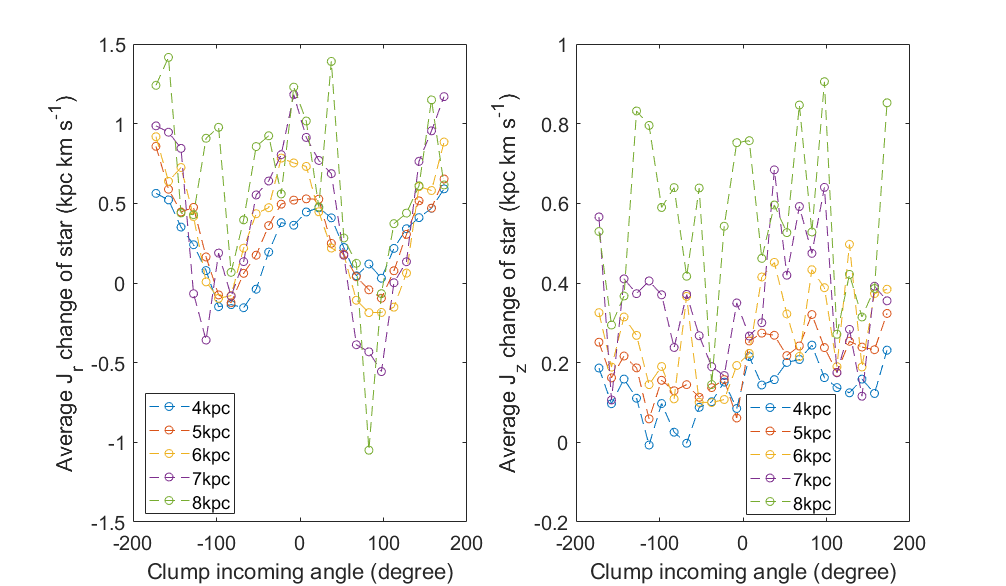}
    \caption{Average $J_r$ and average $J_z$
    changes as a function of clump incoming angles. Colors reflect
    encounters occurring at different galactic radii. The bin width in both panels is 15 degrees.
    }
    \label{fig:ff J_r and angle}
\end{figure*}

Figure~\ref{fig:ff J_r and angle} shows $J_r$ and $J_z$ changes as a function of clump incoming 
angle. The $J_r$ plot reveals a shape of "W" with the troughs around 
-90$^\circ$ and 90$^\circ$. Similar to $L_z$, the $J_r$ changes can be 
estimated by using the trajectory of the star on the disc plane, so they
are related to clump incoming angles. We will see this in detail later.
The $J_z$ changes are positive and show little dependence on clump incoming angle. 
This makes sense because $J_z$ is associated with motions in the $z$-direction and clump incoming angles give no information on that. When examining $J_z$ changes by using motions in the $z$-direction, one will find that the cause of $J_z$ changes shares a similar nature with the $J_r$ changes. We will not specifically discuss $J_z$ evolution in this paper.

\begin{figure}
	\includegraphics[width=\columnwidth]{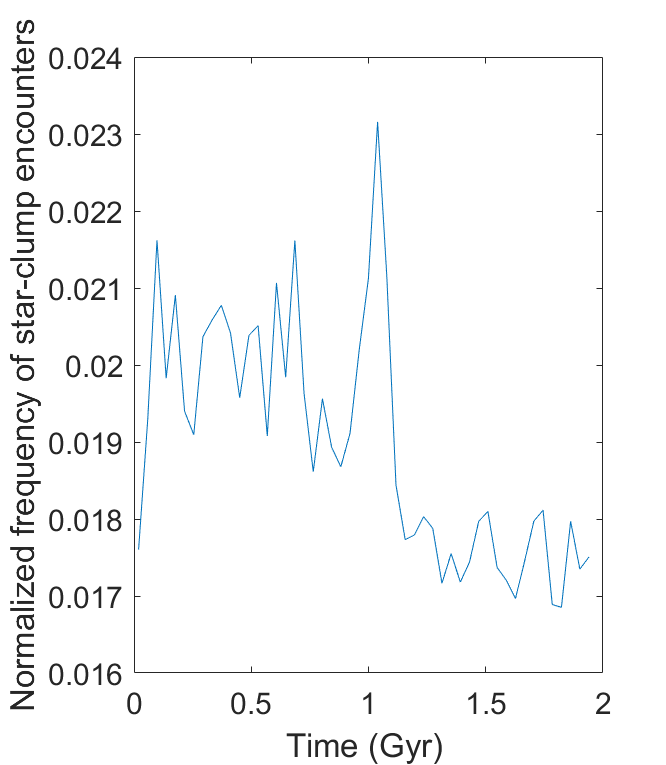}
    \caption{Normalized frequency of star-clump encounters as a function of time from 0 to 1.96 Gyr. Only encounters that occur at a galactic radius greater
    than 2 kpc are counted. The width of a bin is 39.2 Myr. 
    }
    \label{fig:ff collision time}
\end{figure}

Figure~\ref{fig:ff collision time} shows the frequency of star-clump encounters as a function of time.
As mentioned earlier, an encounter involving multiple clumps simultaneously will be 
counted multiple times.
To reduce the effect of multiple counting, we select only encounters occurring at
a galactic radius greater than 2 kpc when making the plot.
In the figure, the normalized encounter frequency has a downward trend, and a large fluctuation 
happens around 1 Gyr. These features match up with what we see in Figure~\ref{fig:ff time distri}. When the
encounter frequency goes down, so does the large $L_z$ change frequency.
This confirms that star-clump encounters generate stellar $L_z$ changes.
The downward trend in Figure~\ref{fig:ff collision time} may be caused by increased random motion in the $z$-direction.
When the stellar disc gets thicker, the separation between stars and clumps in  the $z$-direction 
increases, and the chance of a star entering the three-dimensional neighbourhood of a clump goes down.

\begin{figure*}
	\includegraphics[width=\textwidth]{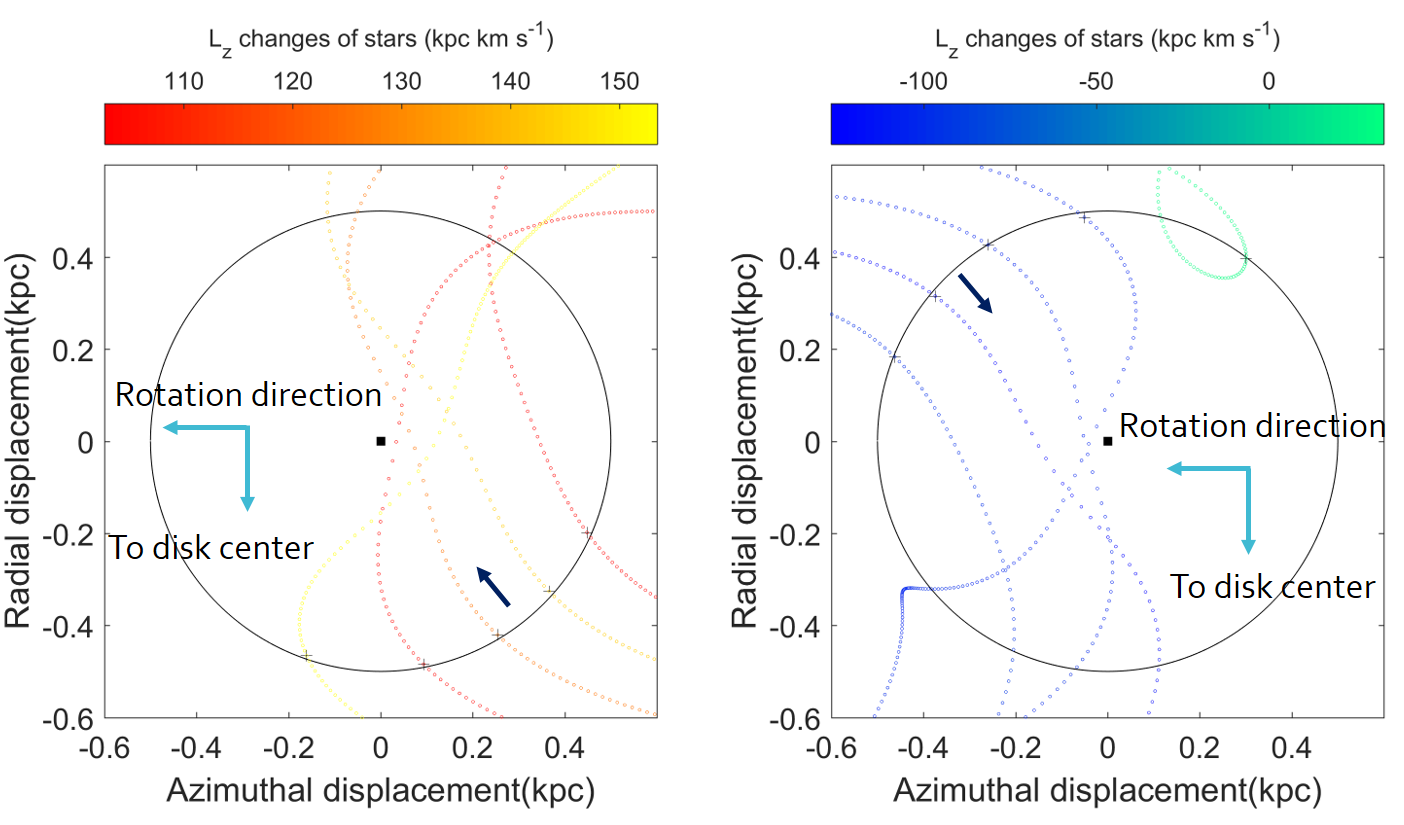}
    \caption{Face-on view of stellar orbits in the rotating frame centred 
    on the clump, produced in the local shearing sheet model. The black
    square in the centre is the position of the clump. Dotted curves are
    positions of the stars observed every 1.96 Myr. A star initially moves in a circular orbit at 8 kpc.  Black plus signs mark
    the positions where the star enters the clump neighbourhood. Black arrows show the direction of stellar motions in this frame.
    On each panel, downwards is the 
direction to the galactic centre, and left is the direction of disc rotation.
The left panel shows the cases where the clump is initially placed ahead of the star in the azimuthal direction by roughly 1 kpc. In this panel, the five stellar tracks ordered by their plus signs from right 
to left correspond to clump galactic radii $R_0$ of 8.4, 8.5, 8.6, 8.7, 8.8 kpc, respectively. 
The right panel shows the cases where the clump is initially placed behind the star in the azimuthal direction by roughly 1 kpc. The five stellar tracks ordered by their plus signs from left
to right correspond to clump galactic radii $R_0$ of 7.6, 7.5, 7.4, 7.3, 7.2 kpc, respectively. 
The color of dotted curves indicates the $L_z$ change of the star, measured by using the values when
the star enters and leaves the clump neighbourhood.
    }
    \label{fig:ff orbit clump frame}
\end{figure*}

\section{Star-clump encounters in the local shearing sheet model}
\label{sec:f_star_clump_enc}

In the previous section, we showed the stellar profile evolution towards an 
exponential, suggested that the large $L_z$ changes are caused by close 
star-clump encounters, and illustrated statistics on action integral changes 
during encounter events. Particularly, we presented the distribution of clump
incoming angles and demonstrated a dependence of $L_z$ and  $J_r$
changes on  these angles. To dig deeper into the shape of the distribution curve
and the explanation of the dependence, we  use the local shearing sheet model. 
As mentioned in Section~\ref{sec: f motion stars rotating frame}, the local shearing sheet model depicts
the trajectory of a star under the influence of a clump on a perfectly
circular orbit, on top of an axisymmetric galaxy potential  with a flat  
rotation curve.

\subsection{Stellar orbits in the clump neighbourhood }
\label{sec:f_stellar_orb}

In this subsection, we try to understand the dependence of $L_z$ and  $J_r$
changes of clump incoming angles. Before investigating this issue,
let us look at stellar orbits during an encounter in the local shearing sheet model. Assume a star is initially in a perfect circular orbit at 
a galactic radius $r_0 = 8$ kpc. We set the rotational velocity $v_0$ in 
the model to be 113 km\,s$^{-1}$, which is the value in the galaxy simulation, so we can compare results with the simulation later. 
A clump is placed about 1 kpc away from the initial position of the star in the azimuthal direction.
The galactic radius $R_0$ of the clump can vary from 7 kpc to 9 kpc with an
increment of 0.1 kpc. 
The galactic radius $R_0$ of the clump affects whether the star can enter
the neighbourhood of the clump at a later time.
Figure~\ref{fig:ff orbit clump frame} shows the stellar orbit in a rotating frame centred on the clump
for the cases where the star goes into the clump neighbourhood. The black square
at the centre of each panel is the position of the clump. The black circles have a radius of 0.5 kpc. Dotted curves are tracks
of the star in the neighbourhood of the individual clump realizations (superposed in a common `clump frame' in the figure). The time difference between two dots
is 1.96 Myr.
Black plus signs on the black circle are locations where the star enters the circle.
From these signs, one can tell the direction of stellar motions.
The color of the dotted curve indicates the $L_z$ change of the star from the circle
entrance point to the circle exit point. In each panel, downwards is the 
direction to the galactic centre, and left is the direction of disc rotation.
The left panel shows the cases where the star is initially placed behind of the clump in the azimuthal direction, so the star tracks come in from the lower right. In this panel, the five stellar tracks ordered by their plus signs from right 
to left correspond to clump galactic radii $R_0$ of 8.4, 8.5, 8.6, 8.7, 8.8 kpc, respectively. Recall that the star is always at 8 kpc radius, so the stellar tracks that start furthest in radius (y-coordinate) from the clump correspond to clump positions furthest from 8 kpc.
Although the coordinate of the clump is (0,0) for all cases  on this
local frame of reference, the site of the clump in the galactic disc is different for each case.
The right panel shows the scenarios where the clump is initially placed behind 
the star in the azimuthal direction. The five stellar tracks ordered by their plus signs from left
to right correspond to clump galactic radii $R_0$ of 7.6, 7.5, 7.4, 7.3, 7.2 kpc,  
respectively.

When the clump has a galactic radius greater than that of the initial star (left panel of Figure~\ref{fig:ff orbit clump frame}),
the star approaches the clump from a lower galactic radius and then gets thrown outwards to a higher radius. The stellar trajectory in the rotating
frame bends before it passes the clump, so the trajectory does not go around the clump.
When the clump has a galactic radius less than the star,
the star meets the clump from a higher galactic radius and then gets sent inwards
to a lower radius in front of the clump with a decrease in the $L_z$, except for $R_0 = 7.2$ kpc  which barely intercepts the clump interaction radius.
These trajectories  from above also do not  go around the clump.
The stellar paths in the figure agree with orbits in Figure 8.12 of \citet{2008book}.
The orbits of stars in \citet{2008book} bend before they pass the mass point when their initial
galactic radii have a small difference from the mass point. When the difference in
galactic radii gets large, Figure 8.12 in \citet{2008book} also shows a trajectory where the star
reverses its motion in the radial direction, as  does
the $R_0 = 7.2$ kpc case in our right panel.

To understand the $L_z$ change of a star during an encounter, we need to figure
out the torque exerted on the star by the gravity of the clump. The magnitude of 
the torque depends on the strength of the gravity and its direction relative to
the disc rotation. To better examine the torque from the clump, the position of the clump during an
encounter is presented in a moving frame centred on the star.  On
the  stellar moving frame,
at each time point, we create a Cartesian coordinate system with the origin at
the position of the star. Its horizontal and vertical axes are along the azimuthal  and radial directions at the star, respectively.

\begin{figure*}
	\includegraphics[width=\textwidth]{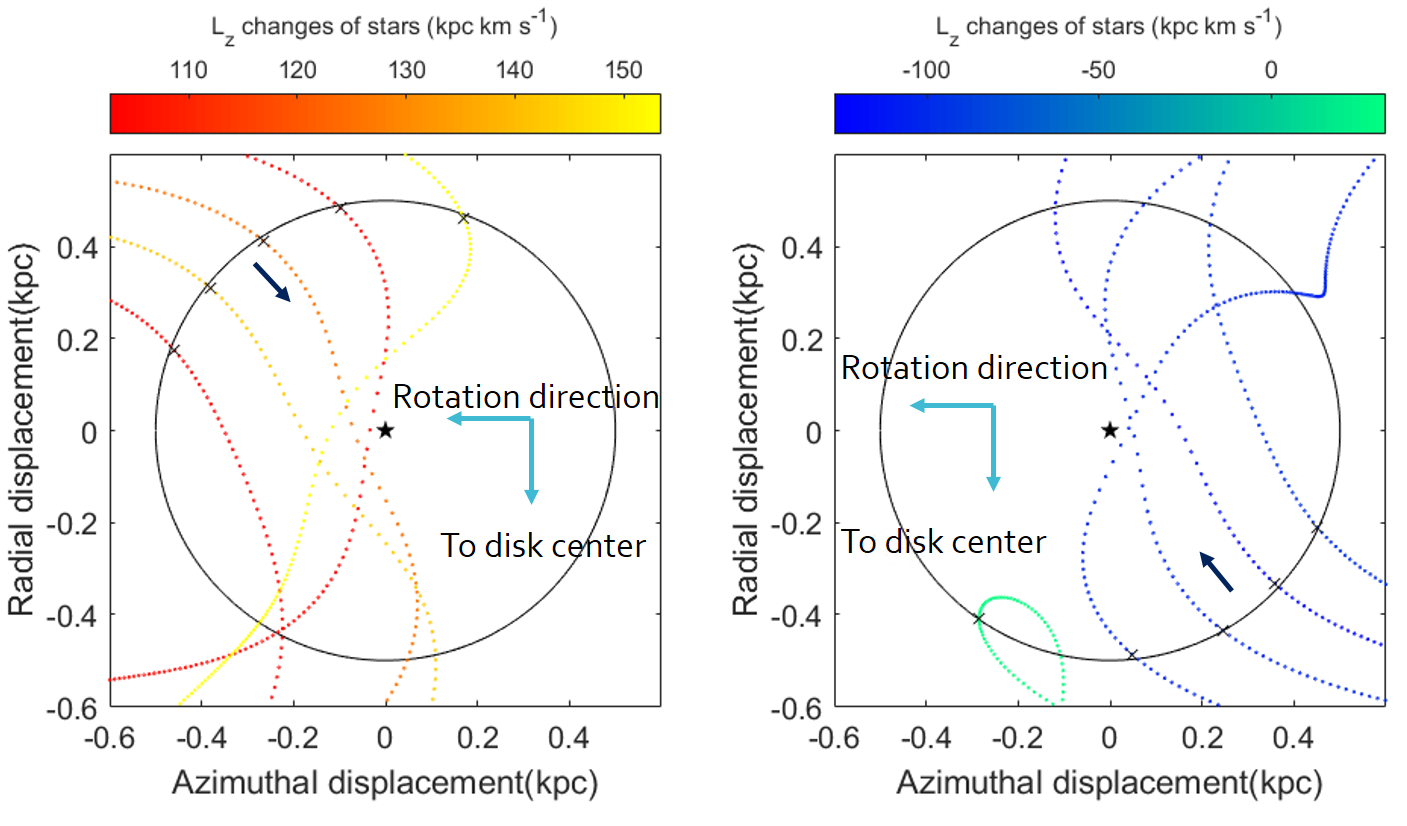}
    \caption{Clump positions in the moving frame centred 
    on the star, shown for cases in Figure~\ref{fig:ff orbit clump frame}. The black
   pentagram at the centre is the position of the star. Dotted curves are
    positions of the clump observed every 1.96 Myr.  Black crosses mark
    the positions where the clump enters the star neighbourhood. Black arrows show the direction of clump motions in this frame.
    On each panel, downwards is the 
direction to the galactic centre, and left is the direction of disc rotation.
On the left panel, the five tracks
 ordered by their entrance points from left
to right correspond to clump galactic radii $R_0$ of 8.4, 8.5, 8.6, 8.7, 8.8 kpc,  respectively.
On the right panel, the five tracks
 ordered by their entrance points from right
to left correspond to clump galactic radii $R_0$ of 7.6, 7.5, 7.4, 7.3, 7.2 kpc,  respectively.
The color of dotted curves indicates the $L_z$ change of the star.
    }
    \label{fig:ff orbit star frame}
\end{figure*}

Figure~\ref{fig:ff orbit star frame} shows the positions of the clump in this moving frame for orbits in Figure~\ref{fig:ff orbit clump frame}.
In Figure~\ref{fig:ff orbit star frame}, the black pentagram at the centre of each panel is the position of 
the star in this local frame.  The black circles have a radius of 0.5 kpc.
In each panel, downwards is the 
direction to the galactic centre, and left is the direction of disc rotation.
 Dotted curves are positions 
of the clump. When a star enters the neighbourhood of the clump, it can be viewed from the perspective of the star as
the clump entering the star neighbourhood. Black crosses mark the entrance points where  
the clump goes into the circle. They are also used to determine the clump incoming angles.
The colors of the dotted curves show the $L_z$ changes of the star as before.

On the left panel, the five tracks
 ordered by their entrance points from left
to right correspond to clump galactic radii $R_0$ of 8.4, 8.5, 8.6, 8.7, 8.8 kpc,  
respectively. The clump incoming angles are 
159.3$^\circ$, 140.9$^\circ$, 122.9$^\circ$, 101.5$^\circ$, and 69.8$^\circ$, respectively.
The $L_z$ changes of the star are 102.7,
142.7, 125.0, 104.0, and 153.5, in units of kpc\,km\,s$^{-1}$.
In this panel, as the clump galactic radius $R_0$ increases, 
the clump incoming angle goes down until the star cannot enter the
clump neighbourhood.
On the right panel, the five tracks
 ordered by their entrance points from right
to left correspond to clump galactic radii $R_0$ of 7.6, 7.5, 7.4, 7.3, 7.2 kpc,  
respectively.
The clump incoming angles are -25.0$^\circ$, -42.9$^\circ$, -60.6$^\circ$, -84.4$^\circ$, and -124.9$^\circ$, respectively.
The $L_z$ changes of the star are
-97.3, -126.8, -105.3, -105.5, and 33.2, in unit of kpc\,km\,s$^{-1}$.
Here, as the clump galactic radius $R_0$ falls, 
the clump incoming angle decreases until the star cannot enter the
clump neighbourhood.

In Figure~\ref{fig:ff orbit star frame}, when the position of the clump is in the second or third
quadrant, the torque from clump gravity on the star is positive. When the clump position is in the first or fourth quadrant, the torque is negative. The magnitude of the torque is higher if the clump has a 
smaller distance to the star or the direction from the star to the clump
is closer to the azimuthal axis. For any orbit in the left panel,
most clump positions are in the second or third quadrant. Since the 
angular moment change is the time integral of the torque, the $L_z$
changes in this panel are all positive.
In the right panel, the track of $R_0 = 7.2$ kpc has clump positions
in the third quadrant, so the $L_z$ change is positive. For the rest of orbits,
most clump positions are in the 
first or fourth quadrant, so their $L_z$
changes are negative.

We can also use clump positions in Figure~\ref{fig:ff orbit star frame} to explain the $J_r$
changes of the star, although  $J_r$ changes are a little more complicated. When the clump is in the first or second quadrant in 
Figure~\ref{fig:ff orbit star frame}, the gravity from the clump on the star has an outward radial component, and this component will generate an outward impulse
on the radial motion of the star as time goes on.
When the clump is in the third or fourth quadrant,  gravity
will produce an inward impulse on the star.
If the outward impulse is greater than the inward impulse, 
the star receives a net outward impulse.
Otherwise, it receives a net inward impulse.
The $J_r$ change of the star depends not only on the direction
of the net impulse received, but also on the phase of the radial motion
when the star enters the clump neighbourhood.
If the star is initially moving outwards in the radial direction or 
has an outwards momentum,
a net outward impulse will increase its radial action and a
net inward impulse will reduce it. If the star is initially moving
inwards, the change of $J_r$ is the opposite. If the star does not
initially have any random motion in the radial direction,
a net impulse regardless of its direction will increase the 
$J_r$ of the star. 
In Figure~\ref{fig:ff orbit star frame}, the track with a clump incoming angle of 101.5$^\circ$
is roughly symmetric about the azimuthal axis across the star. 
The magnitude of the outward impulse due to clump positions in the first and second
quadrants is close to the magnitude of the inward impulse from the third and fourth
quadrants. Therefore, the net impulse is very small, which means 
the $J_r$ change should also be small. 
The track with an incoming angle of 159.3$^\circ$
has most of the clump positions in the third quadrant.
The net impulse should be relatively large, since most of the clump positions
contribute to the impulse in the same radial direction.
In the simulated galaxy, the radial velocity 
dispersion of stars is much less than the rotational velocity in the outer disc. Usually we can assume that the radial actions of stars before encounter events are negligible there. Therefore, a large net impulse at the angle of 159.3$^\circ$ indicates a big $J_r$ change.
These results agree with the $J_r$ change curve in Figure~\ref{fig:ff J_r and angle}.
In Figure~\ref{fig:ff J_r and angle}, the $J_r$ change at the angle of 101.5$^\circ$ is close
to 0, and the change at 159.3$^\circ$ is around the peak value.

\begin{figure*}
	\includegraphics[width=\textwidth]{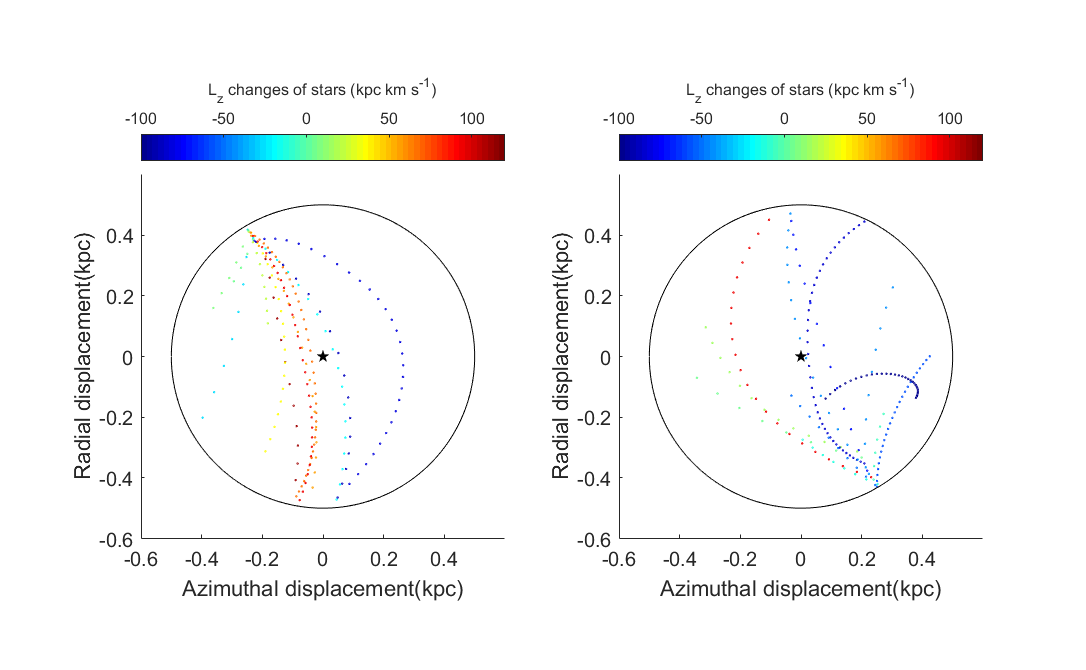}
    \caption{Face-on view of clump positions in the moving frame centred 
    on the star, shown for encounters occurring at 8 kpc in the simulated galaxy with clump incoming angles around 120$^\circ$ and -60$^\circ$. The black
   pentagram at the centre is the position of the star. Dotted curves are
    positions of the clump observed every 1.96 Myr. 
    On each panel, downwards is the 
direction to the galactic centre, and left is the direction of disc rotation.
 The left panel and the right panel show clump incoming angles of 120$^\circ$ and -60$^\circ$, respectively. 
The color of dotted curves indicates the $L_z$ change of the star.
    }
    \label{fig:ff L gadget 120}
\end{figure*}

\begin{figure*}
	\includegraphics[width=\textwidth]{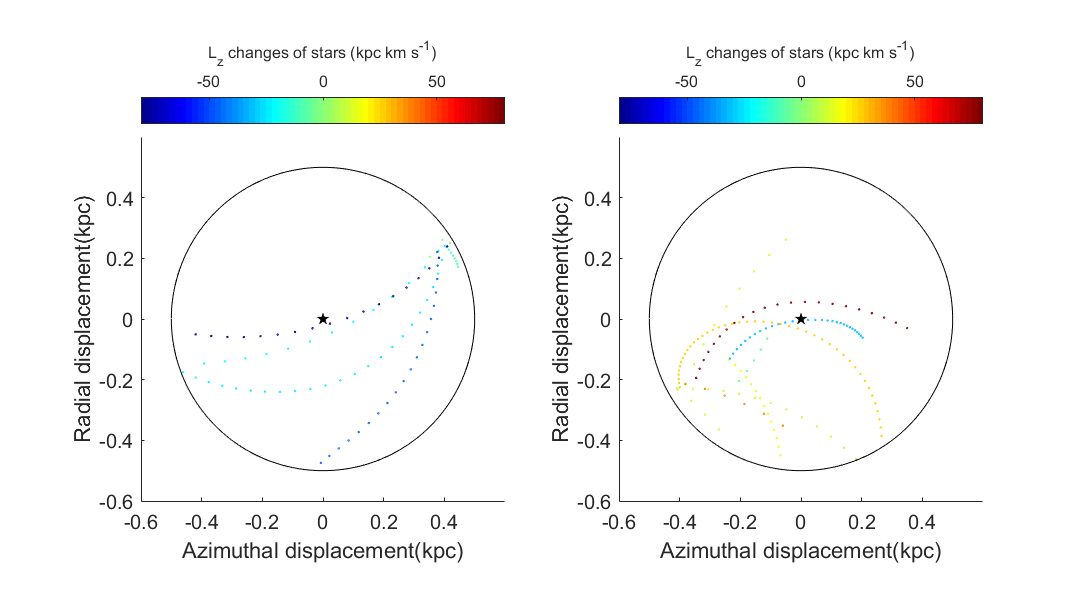}
    \caption{Face-on view of clump positions in the moving frame centred 
    on the star, shown for encounters occurring at 8 kpc in the simulated galaxy with clump incoming angles around 30$^\circ$ (left panel) and -150$^\circ$ (right panel). 
    }
    \label{fig:ff L gadget 30}
\end{figure*}

\begin{figure*}
	\includegraphics[width=\textwidth]{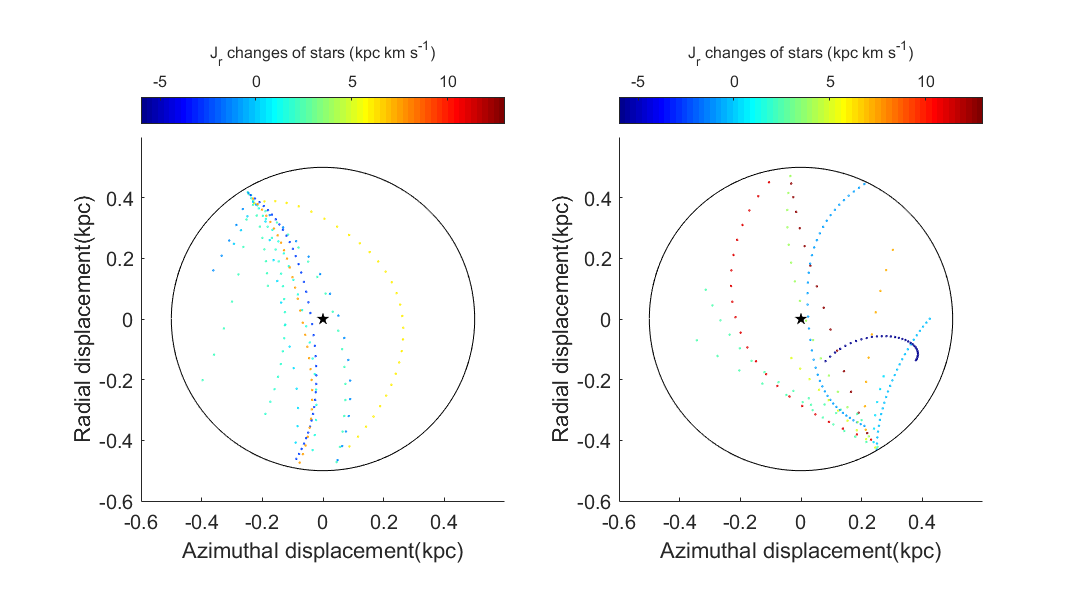}
    \caption{$J_r$ changes for trajectories in Figure~\ref{fig:ff L gadget 120}.
 The left panel and the right panel show clump incoming angles of 120$^\circ$ and -60$^\circ$, respectively. 
The color of dotted curves indicates the $J_r$ change of the star.
    }
    \label{fig:ff jr gadget 120}
\end{figure*}

\begin{figure*}
	\includegraphics[width=\textwidth]{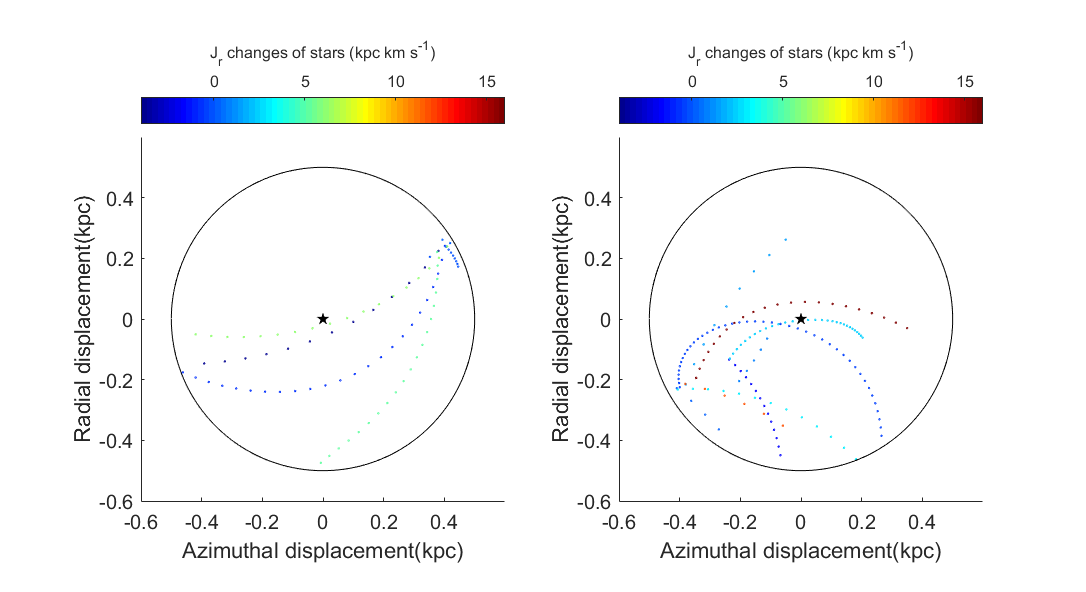}
    \caption{$J_r$ changes for trajectories in Figure~\ref{fig:ff L gadget 30}.
 The left panel and the right panel show clump incoming angles of 30$^\circ$ and -150$^\circ$, respectively. 
The color of dotted curves indicates the $J_r$ change of the star.
    }
    \label{fig:ff jr gadget 30}
\end{figure*}
Now let us look at encounter orbits in the simulated disc at the 
galactic radius of 8 kpc. There are two major differences between 
the simulated disc and the local shearing sheet model. First,
neither clumps nor stars in the galaxy simulation move in perfect 
circles. Their orbits can carry random energy in the radial or azimuthal
direction. Second, clumps and stars can have velocity components
perpendicular to the galaxy midplane. When a star meets a clump, 
their $z$-coordinates may not be the same.

Figure~\ref{fig:ff L gadget 120} shows the clump paths in the moving frame of the star
for encounters at 8 kpc in the simulated galaxy. The left panel
shows encounters with the clump incoming angle of 120$^\circ$
and the right panel shows encounters with the incoming angle
of -60$^\circ$. Again,  downwards is the 
direction to the galactic centre, and left is the direction of disc rotation. The time difference between two successive dots on a track
is 1.96 Myr, the same as the value used in the local shearing sheet model. The color of a track tells the $L_z$ change of the star during
the encounter. As one can see, some clump tracks do not start or end
at the boundary of the 0.5 kpc circle. This is because the clump and 
the star do not have the same $z$-coordinate at the moment when the star enters
or leaves the clump neighbourhood  and the projected position of the star's impact on the neighbourhood sphere is inside the sphere's projected circle.

In the left panel  of Figure~\ref{fig:ff L gadget 120}, most of the clump paths do not  go around the star at
the centre. This is what the local shearing sheet model predicts. However, three curves do go around the star before they reverse their motion
in the radial direction, due to  non-zero random  motions. The $L_z$ changes for these paths are  
 negative, which is the opposite of the other paths.
This means that the torque from the clump at the closest approach
dominates the  change in  $L_z$ of the star. When the clump path envelops the star,
the clump position at the closest approach is in the first or  fourth
quadrant. The negative torque from the clump at this moment is large,
because the gravity of the clump reaches its peak value at this  point. Usually, this negative torque is large enough to make the
whole time integral of the torque negative,  so the $L_z$ change becomes
negative. When calculating the average $L_z$ changes for all orbits in this panel, the value is positive, as most clump tracks do not  go around
the star. As we see in Figure~\ref{fig:ff L_z and angle}, the average $L_z$ change at the
clump incoming angle of 120$^\circ$ is indeed positive.

Clump paths in the right panel of Figure~\ref{fig:ff L gadget 120} give similar results.
Most tracks are in the first and  fourth quadrants, in accordance with
the local shearing sheet model.
When the clump path envelops the star, the $L_z$ change switches from
negative to positive. Since most of the tracks do not envelop the star,
the average $L_z$ change for all orbits is negative. This matches
the $L_z$ change value at -60$^\circ$ on the right panel of Figure~\ref{fig:ff L_z and angle}.

The left and the right panels of Figure~\ref{fig:ff L gadget 30} show the clump paths and the $L_z$ changes
for encounters with clump incoming angles of 30$^\circ$ and -150$^\circ$, respectively,  in the simulated galaxy. 
Since these two clump incoming angles are in the forbidden intervals,
the number of orbits in this figure is less than that in Figure~\ref{fig:ff L gadget 120}.
On the left panel, almost all the clump tracks go around the star, which
is different from what we see in Figure~\ref{fig:ff L gadget 120}. On the right panel,
the dark red curve above the star and the yellow curve under the star
both have  positive $L_z$ changes. It seems that whether or not a curve
 goes around a star does not affect the sign of $L_z$ change here.
Another difference is that
the clump position at the closest approach in Figure~\ref{fig:ff L gadget 30} usually has a direction closer to the radial direction, rather than the azimuthal direction. This makes the torque at the closest approach weaker.   
As a result, the magnitude of $L_z$ change for each orbit is less than
that in Figure~\ref{fig:ff L gadget 120} on average. 
On the left panel, the closest approaches of most clump tracks are
in the right semicircle, so the average $L_z$ change at the angle of
30$^\circ$ is negative. 
On the right panel, the closest approaches of most clump tracks are
in the left semicircle, so the average $L_z$ change at the angle of
-150$^\circ$ is positive.
These results agree with the average $L_z$ change curves in Figure~\ref{fig:ff L_z and angle}.

Figure~\ref{fig:ff jr gadget 120} shows the $J_r$ changes for the encounter events in Figure~\ref{fig:ff L gadget 120}. 
Since the $J_r$ changes depend not only on the net radial impulse from 
the clump but also on the direction of initial radial
momentum of the star, orbits with similar clump tracks can
have different $J_r$ changes. For example, the orange curve and 
the dark blue curve on the left panel overlap, but they have totally
different $J_r$ changes. Also on the left panel, whether or not 
a curve envelops  the star does not seem to affect the $J_r$ change.
In the figure, orbits with a large $J_r$ change are not
symmetric about the azimuthal axis across the star. This aligns with
our earlier argument during the discussion on the $J_r$ changes in the local shearing sheet model.

Figure~\ref{fig:ff jr gadget 30} shows the $J_r$ changes for the encounters at the angle of 
30$^\circ$ and -150$^\circ$. In this figure, curves are far from
symmetric about the azimuthal axis, as most clump positions
are in the bottom semicircle. Therefore, the star receives a relatively
large net radial impulse, indicating a large $J_r$ change.
As shown on the $J_r$ curve on Figure, the average $J_r$ changes at 
the angles of 30$^\circ$ and -150$^\circ$ are greater than those at 
120$^\circ$ and -60$^\circ$.

In summary, the clump incoming angle determines where the clump path
is located in the 0.5 kpc circle. The path tells the magnitude and the direction of the 
stellar acceleration due to the gravity of the clump. 
When the clump incoming angle is 0$^\circ$, the distribution of clump paths 
has most of the clump positions in the right semicircle as all paths start from the right end of the 0.5 kpc circle. This gives a large negative average $L_z$
change. When the clump incoming angle is 180$^\circ$, the opposite is true and this case corresponds to a large positive average $L_z$
change. Individual clump paths in  the 0$^\circ$ and 180$^\circ$ cases  are far from symmetric about the azimuthal axis. Therefore, both cases have a large $J_r$
change. 
When the clump incoming angle is close to 90$^\circ$ or -90$^\circ$,
the $L_z$ change of  an individual path usually has a large magnitude. 
However, as the sign of the $L_z$ change depends on whether the path  goes around the star and both types have similar  probability of occurrence, the average $L_z$
change is close to 0. Because many clump paths are roughly symmetric about the azimuthal axis, the average $J_r$ changes at 90$^\circ$ and -90$^\circ$ are close to 0 as well. 
The above analysis explains the general features
 shown on the $L_z$ curve in Figure~\ref{fig:ff L_z and angle} and the $J_r$ curve in Figure~\ref{fig:ff J_r and angle}.
Angles corresponding to a large average $L_z$ change, regardless of the sign, also tend to give a large average $J_r$ change.

\begin{figure*}
	\includegraphics[width=\textwidth]{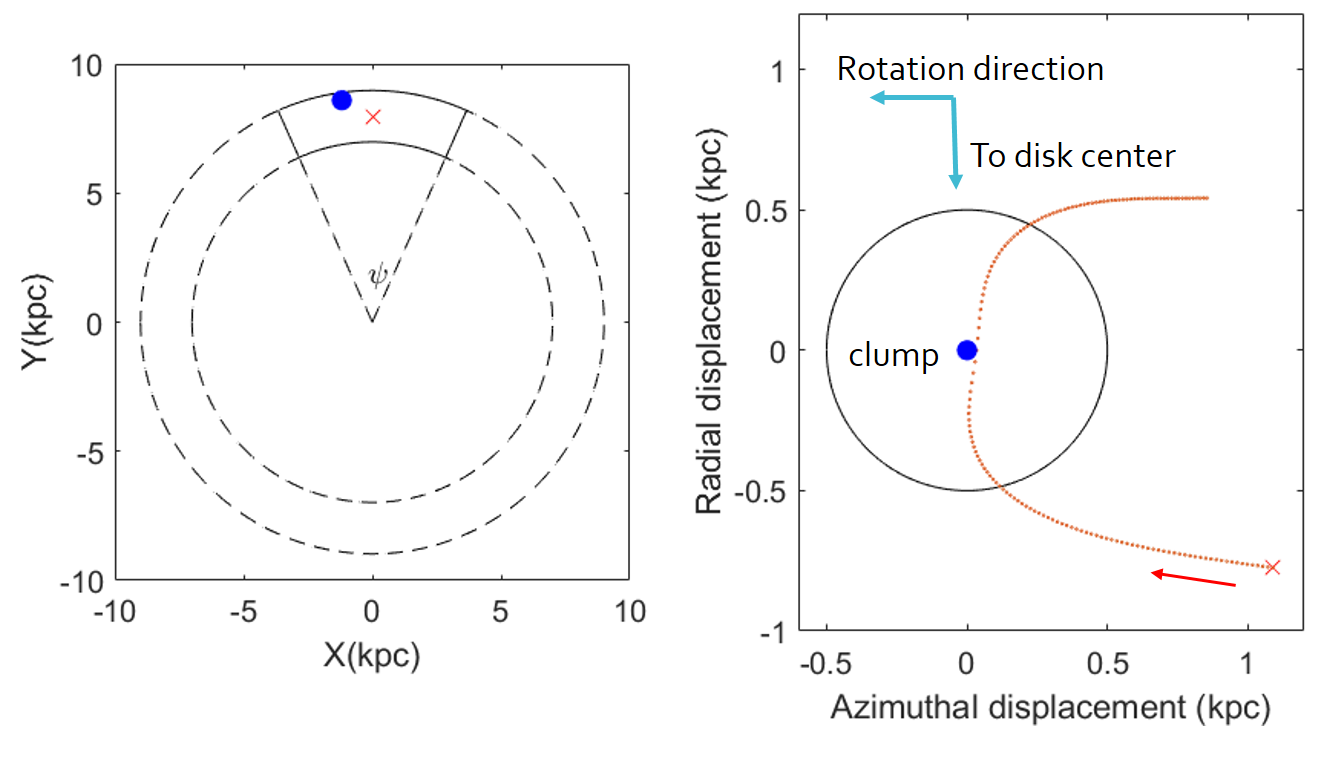}
    \caption{Demonstration of the initial condition setup in the local shearing sheet model. The left panel is the face-on view of the galaxy with
    the galactic centre at the origin. In the model, the initial position of the star, represented by 
    the red cross, is placed at (0, 8). The initial position of the
    clump is constrained in the annular sector with the angle $\psi$ of 48$^\circ$ and the two arcs at 7 kpc and 9 kpc. After the initial position
    of the clump, represented by the blue dot, is determined on the left panel, we calculate
    the orbits of the star in the rotating frame centred on the clump, shown as the dotted curve on the right panel. On the
    right panel, the horizontal axis and the vertical axis represent azimuthal displacement and radial displacement,   respectively. 
    Downwards is the 
direction to the galactic centre, and left is the direction of disc rotation, which is the same as Figure~\ref{fig:ff orbit clump frame}. The red arrow shows the direction of the stellar motion in this frame. 
    }
    \label{fig:ff demo2}
\end{figure*}

As an aside we note that these changes in star-clump collisions are not captured by the impulse approximation. Using the impact parameter and relative velocities of the each collision we computed the change in velocity components predicted by that approximation. We found no significant correlation with the calculated velocity changes.
To illustrate this point, it is worth revisiting the left panel of 
Figure~\ref{fig:ff orbit clump frame}
, which shows a few stellar trajectories relative to the clump. By looking at the space between adjacent dots, one can find that the difference of the in and out speed at the black circle depends on trajectories and can be positive, negative, or almost 0. 
This result is inconsistent with the impulse approximation. The looping orbits relative to the clump show that Coriolis forces are important, and therefore the galactic potential is making the impulse approximation oversimplified.


\subsection{Dependence of \texorpdfstring{$L_z$}{Lg} changes on clump density distribution}
\label{sec:f_depend_lz}

So far, we have provided explanations for the shape of curves on the right panel of Figure~\ref{fig:ff L_z and angle}
and the left panel of Figure~\ref{fig:ff J_r and angle}.
In this subsection, we study the influence of clump density profile on the clump incoming angle distribution, which is
 shown on the left panel of Figure~\ref{fig:ff L_z and angle}. Other factors, such as velocity 
dispersion, azimuthal drift of stars due to relatively high velocity dispersion,
etc., can also affect the distribution of clump incoming angles. Here, we 
constrain ourselves to clump density profile. 

We use the local shearing sheet model to investigate star-clump encounters under different clump surface distributions. 
We choose  the clump density profile to have the following power law surface density distribution
\begin{equation}
    \Sigma_{\text{clump}} (r) \propto r^\alpha ,
	\label{eq:clump density dist}
\end{equation}
where the power $\alpha$ is a negative constant. Notice that the number of 
clumps $n_{\text{clump}} (r)$ in the radial interval [$r$, $r$+d$r$] satisfies the following
\begin{equation}
    n_{\text{clump}} (r) \propto \Sigma_{\text{clump}} (r) r \text{d}r
   \propto r^{\alpha +1} \text{d}r.
	\label{eq:clump number dist}
\end{equation}

Figure~\ref{fig:ff demo2} illustrates how we set up the initial conditions for the 
local shearing sheet model in this case. A star is initially placed at the galactic radius of 8 kpc,
shown as the red cross on the left panel. The azimuthal component of the initial velocity of the 
star follows a normal distribution with a mean of 113 km\,s$^{-1}$ and a
standard deviation of 13 km\,s$^{-1}$. The radial component of the initial velocity follows a normal distribution with a mean of 0 km\,s$^{-1}$ and a
standard deviation of 13 km\,s$^{-1}$. As we mentioned before,  113 km\,s$^{-1}$ is the mean rotational speed of stars at 8 kpc in the N-body simulation. 
13 km\,s$^{-1}$ is the radial velocity dispersion at 8 kpc in the simulation.
A clump is initially placed in an annular sector, shown by the solid lines
on the left panel. The azimuthal span $\psi$ of the annular sector is 48$^\circ$, with the star at the centre. The radial interval of the annular
sector is from 7 kpc to 9 kpc. In this annular sector, the initial position
of the clump follows a uniform distribution in the azimuthal direction and 
a power law distribution described by Equation~\ref{eq:clump number dist} in the radial direction.
The velocity of the clump is always 113 km\,s$^{-1}$ along the azimuthal direction.

After  an initial velocity of the star and an initial position of the clump are determined, the motion of the star
in the rotating frame centred on the clump is calculated by the local shearing sheet model. For example, if the initial velocity of the star
is  113 km\,s$^{-1}$ along the azimuthal direction and the initial position of 
the clump is at the blue dot on the left panel, the path of the star in the
clump frame will be the orange curve in the right panel. Here, the blue dot
represents the position of the clump. The red cross is the initial position
of the star. The horizontal and vertical axis stand for azimuthal displacement and radial displacement, respectively. In this panel, downwards is the 
direction to the galactic centre, and left is the direction of disc rotation.
To obtain the clump incoming angle, one needs to view the orbit
as the trajectory of the clump in the moving frame centred on the star.
Since this conversion has been shown in the previous subsection, we do not present it here.

\begin{table}
	\centering
	\caption{Mean stellar $L_z$ changes during encounters and the frequency ratio of 
	negative clump incoming angles to positive ones under clump density profiles
	with different power $\alpha$ in Equation~\ref{eq:clump density dist}. Encounters
	are generated using the local shearing sheet model at the galactic radius of 8 kpc.
}
	\label{tab:Lch clump distribution table}
	\begin{tabular}{ccc} 
		\hline
		Power $\alpha$ & Mean $L_z$ change  & $n_-$/$n_+$\\
&	(kpc\,km\,s$^{-1}$) &\\
		\hline
			-0.5 & 2.99 $\pm$ 0.49 & 0.919\\
	-1.5 & 0.94 $\pm$ 0.49 & 0.964\\
		-2.5 & -0.03 $\pm$ 0.48 & 1.051\\
		-3.5 & -0.66 $\pm$ 0.49 & 1.108\\
	-4.5 & -0.84 $\pm$ 0.49 & 1.129\\
		-5.5 & -1.62 $\pm$ 0.49 & 1.215\\
			-6.5 & -2.64 $\pm$ 0.49 & 1.231 \\
			
		\hline
	\end{tabular}
\end{table}

To figure out the distribution of clump incoming angles, we run the local shearing sheet model 60 000 times for each clump surface density distribution.
Among the 60 000 samples, we only count those where the star can enter the
clump neighbourhood within 1 Gyr. 
Let $n_+$ and $n_-$ denote the number of encounters with positive 
clump incoming angles and negative clump incoming angles, respectively.
Table~\ref{tab:Lch clump distribution table} shows the mean $L_z$ changes and the $n_-$/$n_+$ ratios for 
clump distributions with different power $\alpha$. As one can see,
the mean $L_z$ change decreases and the $n_-$/$n_+$ ratio increases as the power
$\alpha$ in Equation~\ref{eq:clump density dist} goes from -0.5 to -6.5. When the power $\alpha$ goes down, 
the clump surface density function favours small radii to a larger degree, and more clumps get distributed closer to the inner arc of the annular sector.
As more clumps are initially placed at a galactic radius less than the star, the
star is more likely to approach the clump from a higher orbit, which leads to an
encounter with a negative clump incoming angle. Therefore, the $n_-$/$n_+$ ratio 
increases. From Figure~\ref{fig:ff orbit star frame}, we know that a positive clump incoming angle usually gives
a positive $L_z$ change and a negative angle usually gives a negative $L_z$ change.
As the $n_-$/$n_+$ ratio 
increases, it is natural to see that the mean $L_z$ change falls.

\begin{figure*}
	\includegraphics[width=\textwidth]{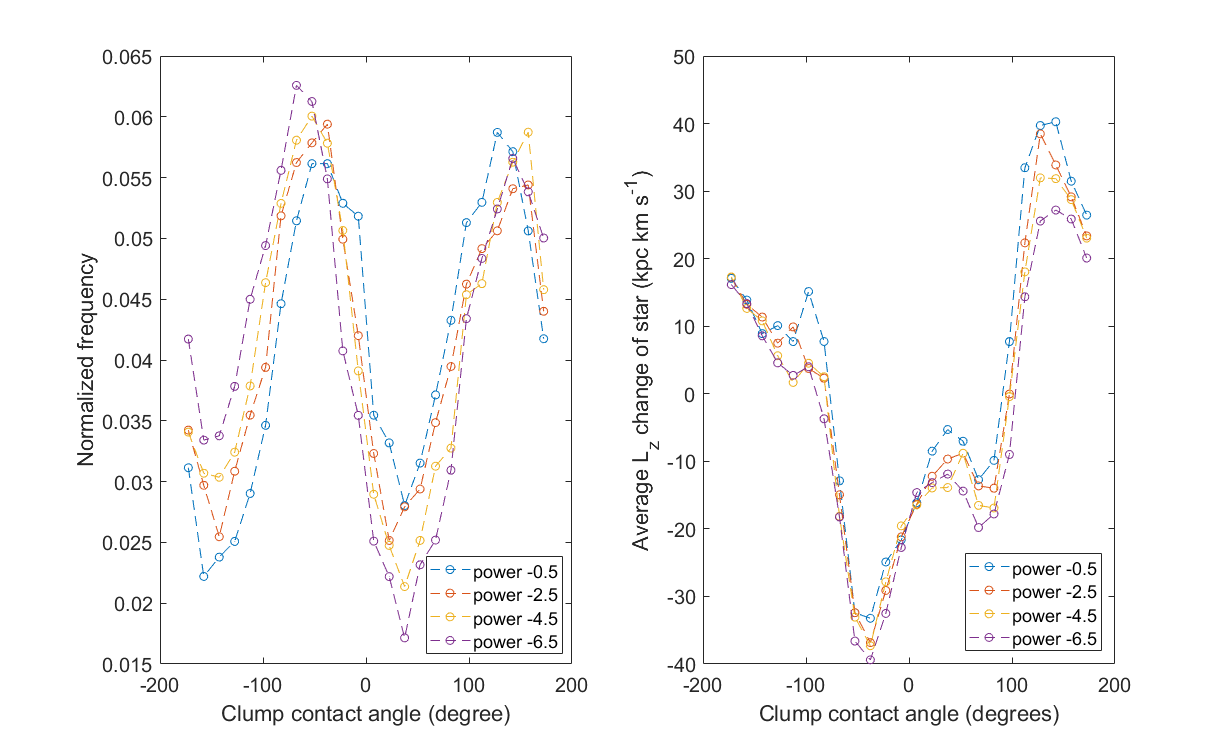}
    \caption{The distribution of clump incoming angles and average $L_z$
    changes as a function of clump incoming angles for clump density
    profiles with different power $\alpha$. Encounters
	are generated using the local shearing sheet model at the galactic radius of 8 kpc. The bin width in both panels is 15 degrees.
        }
    \label{fig:ff loop power}
\end{figure*}

Figure~\ref{fig:ff loop power} shows the clump contact angle distribution and the average $L_z$ change in
each angle bin for several clump surface density profiles. If the initial velocity
of the star does not carry any random orbital energy, the regular intervals of
clump incoming angles are [70$^\circ$, 160$^\circ$] and  [-125$^\circ$, -25$^\circ$] as shown on Figure~\ref{fig:ff orbit star frame}, corresponding to two types of cases where 
the star approaches the clump from a  smaller galactic radius and a greater galactic radius, respectively. On the left panel of Figure~\ref{fig:ff loop power}, 
the frequencies in the first regular interval generally go up and the ones
in the second go down as the power $\alpha$ decreases. This confirms the explanation
given in the previous paragraph. When looking at the two local maximum points of each
distribution curve, the left local maximum frequency gradually rises and the right one declines as the power goes down. The left peak is shorter than the right peak at the power of -0.5, and the opposite holds at the power of -6.5.

As mentioned before, the initial clump surface density in the simulated galaxy follows $e^{-r^2/h^2}$
with a scale length of $h = 5.4$ kpc. The ratio of the density at 7 kpc to the density
at 9 kpc is 3.00, which corresponds to a power $\alpha$ of -4.37. This power will give
a mean $L_z$ change of -0.82 kpc\,km\,s$^{-1}$ when using values at $\alpha$ = -3.5 and $\alpha$ = -4.5 in Table~\ref{tab:Lch clump distribution table} with a linear interpolation.
If we compare it to
the mean $L_z$ change at 8 kpc in Table~\ref{tab:Lch table}, the two numbers are compatible considering 
the uncertainties. When comparing the incoming angle frequency curve at 8 kpc in Figure~\ref{fig:ff L_z and angle} to the frequency curve at $\alpha$ =  -4.5 in Figure~\ref{fig:ff loop power}, they do not match very well.
Besides the difference in clump surface density profiles, the curve of the simulated galaxy is under the influence of stellar motions in the $z$-direction, and the
azimuthal drift due to stellar velocity dispersion makes stars have a smaller mean rotational speed than the clumps. These may be the sources of the divergence.

The result presented so far on the dependence of  $L_z$ changes on the clump density distribution is obtained from the local shearing sheet model. To verify this dependence,
we run another  {\small GADGET}-2 simulation where the initial clump surface density follows a power law with $\alpha$ = -0.5 and all the other parameters are the same as those in the middle column of Table~\ref{tab:para table}.
Table~\ref{tab:Lch table outwards} shows the mean $L_z$ change of a star when it encounters with a clump in this new simulation. One can see that the mean $L_z$ change is positive for $r$ > 5 kpc, so the change in the clump profile flips the sign of $L_z$ changes. At $r$ = 8 kpc, the mean $L_z$ change is 2.89 $\pm$ 0.32 kpc\,km\,s$^{-1}$, which agrees with the value for $\alpha$ =  -0.5 in Table~\ref{tab:Lch clump distribution table}. As for $r \leq$  5 kpc, the  mean $L_z$ changes there are close to values in Tables~\ref{tab:Lch table}. This is mainly because the clump density gradient at $r \leq$  5 kpc in the original simulation is not far away from a power law with $\alpha$ = -0.5. For example, with an $e^{-r^2/h^2}$ clump profile and $h = 5.4$ kpc, the ratio of the clump density at 2 kpc to the density
at 4 kpc in the original simulation is 1.51. In the new simulation with a 1/$\sqrt{r}$ clump profile, this ratio is 1.41, very close to 1.51. 


\begin{table}
	\centering
	\caption{The mean $L_z$ change of  a star during one encounter with a clump at different Galactic radii in the  {\small GADGET}-2 run starting with a 1/$\sqrt{r}$ clump surface density. The $L_z$ change is calculated using the $L_z$ value when a star enters a circle of 0.5 kpc radius around a clump and the one when the star goes out of the circle. 
}
	\label{tab:Lch table outwards}
	\begin{tabular}{ccc} 
		\hline
		Galactic radius & Mean $L_z$ change &   Average number of encounters \\
(kpc)&	(kpc\,km\,s$^{-1}$)  &   per star in one gigayear\\
		\hline
			1 & 0.07 $\pm$ 0.01 & 46.2\\
	2 & -0.31 $\pm$ 0.03 &  28.7\\
		3 & -0.45 $\pm$ 0.05 & 17.8\\
		4 & -0.08 $\pm$ 0.08&14.3 \\
	5 & -0.25 $\pm$ 0.10&  10.7\\
		6 & 1.98 $\pm$ 0.17 & 11.2\\
			7 & 1.49 $\pm$ 0.24 & 10.1\\
				8 & 2.89 $\pm$ 0.32  & 8.5\\
				9 &  4.12 $\pm$ 0.49 & 5.3\\
					10 &  1.49 $\pm$ 0.99 & 2.3 \\
		\hline
	\end{tabular}
\end{table}

\begin{figure*}
	\includegraphics[width=\textwidth]{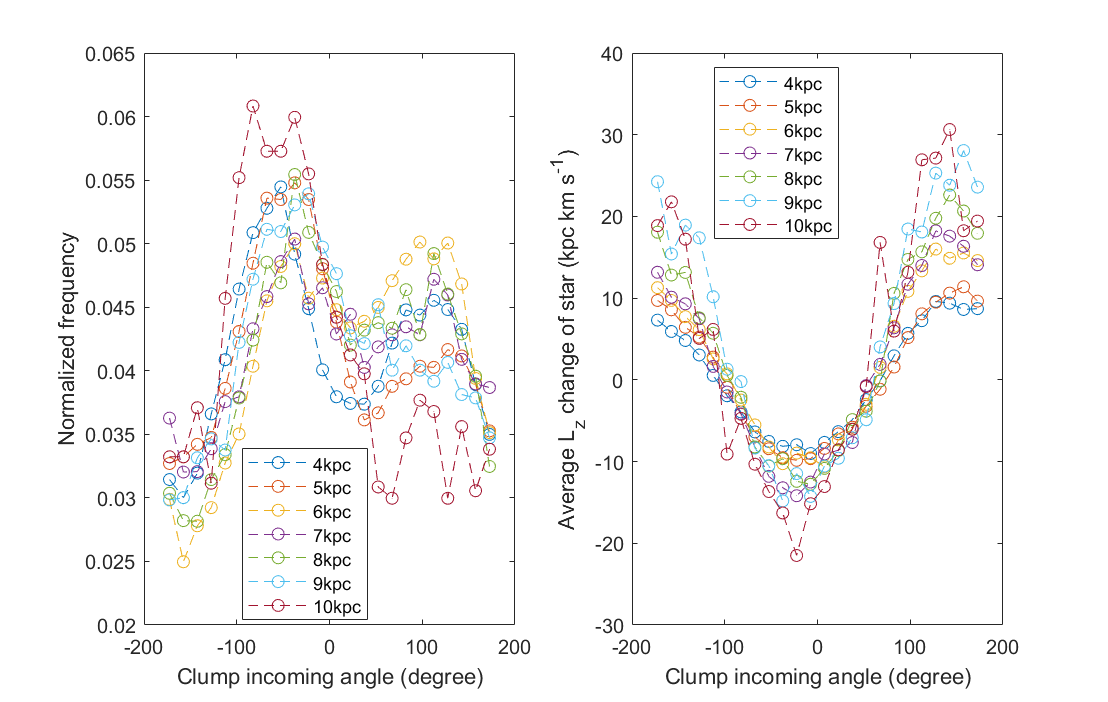}
    \caption{The distribution of clump incoming angles and average $L_z$
    changes as a function of clump incoming angles in the {\small GADGET}-2 run starting with a 1/$\sqrt{r}$ clump surface density. Colors reflect
    encounters occurring at different galactic radii. The bin width in both panels is 15 degrees.
    }
    \label{fig:ff L_z and angle outwards}
\end{figure*}

Figure~\ref{fig:ff L_z and angle outwards} shows the distributions of clump incoming angles and the dependence of $L_z$ changes on the angles in the new simulations. When comparing the left panel with that of Figure~\ref{fig:ff L_z and angle}, one can find that the new simulation has a larger frequency in the angle interval of [70$^\circ$, 160$^\circ$]. As this angle range corresponds to events where a star meets a clump that has a slightly larger orbital radius than the star,  this confirms the result that a shallower clump gradient makes stars more likely to approach a clump from a lower orbit, which has been shown on Figure~\ref{fig:ff loop power}.
Overall, results from the new simulation are consistent with the findings of   the shearing sheet model.

\begin{figure}
	\includegraphics[width=\columnwidth]{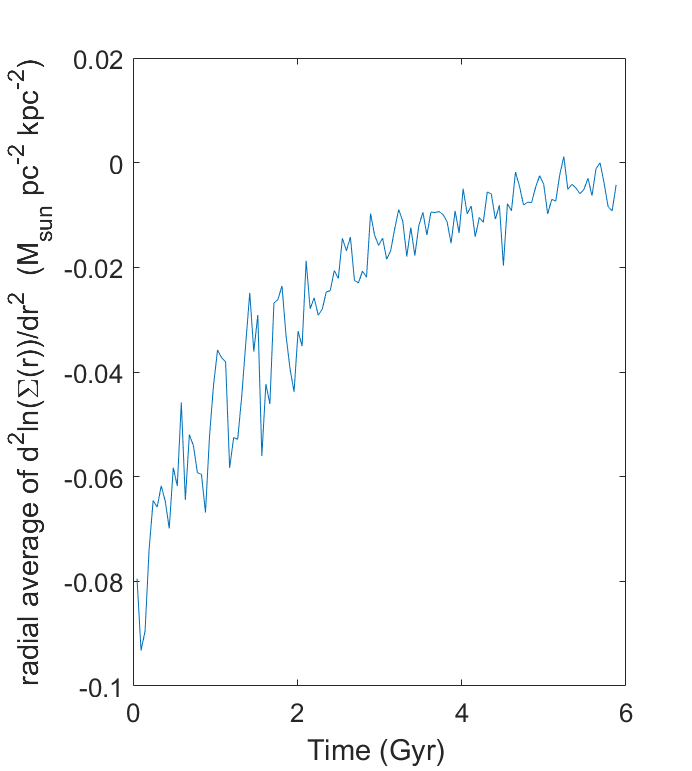}
    \caption{Time evolution of the second derivative of the logarithm radial stellar profile in the {\small GADGET}-2 run starting with a 1/$\sqrt{r}$ clump surface density. 
The radial average is taken from 2 kpc to 9 kpc.
    }
    \label{fig:ff sec deriv new run}
\end{figure}

Although the initial clump profile gets changed in the new simulation, the stellar surface density profile still evolves from $e^{-r^2/h^2}$ to an exponential as the original simulation. Figure~\ref{fig:ff sec deriv new run} shows the second derivative
of the logarithm of stellar surface density as a function of time. Similar to the curve in Figure~\ref{fig:ff sec deriv}, the second derivative gradually increases from a negative value to zero, which indicates an exponential profile. Even though the $L_z$ changes in the outer disc show an outwards bias, this run still produces an exponential stellar profile. An inwards scattering bias is not a prerequisite for the formation of an exponential.

\section{Discussion and Conclusions}
\label{sec:f_discuss_conclu}

The analysis of a self-gravitating disc shows that
star-clump encounters in a galactic disc where hundreds of massive objects exist can contribute to more than 95\% of large stellar angular momentum changes, which drive the surface density profile to evolve towards an exponential shape.
The rate of profile evolution has a positive correlation with the frequency of star-clump
encounters. In the simulated galaxy, both the evolution rate and the
encounter frequency tend to decline with time, as the stellar disc thickens and clumps fall into the galactic centre.

Besides angular momentum, star-clump encounters can also alter 
the radial action and the vertical action of the involved star.
An average encounter event increases the star's radial action and 
vertical action, leading to manifest disc heating.  Velocity changes in the scattering events are not well approximated by the classical Impulse Approximation. These results generally agree with those of
\citet{Spitzer1953} and \citet{Hanninen2002MNRAS.337..731H}. 
They differ somewhat from the results of \citet{D'Onghia2013ApJ...766...34D} and \citet{Vera-Ciro2016} which find little vertical heating. The more massive clumps in the present models scatter more strongly and affect stars at greater distances. Moreover, these more massive clumps move radially through the stellar disc and stellar orbits get more eccentric after initial scattering events, with both effects enhancing scattering.

Detailed analysis on stellar orbits
shows that the angular momentum change and the radial action change during
an encounter depend on the direction from which a star approaches a clump.
A star that enters the neighbourhood of a clump from a greater galactic radius is more likely to lose angular momentum, and one that goes into the clump neighbourhood from a smaller radius has more chance to gain.
The radial action change is more pronounced when a star 
approaches a clump along the azimuthal direction, and less when the
approach is along the radial direction. 

The mean angular momentum change of stars during encounter events
can be affected by the surface density distribution of the clumps.
In a region where the clump density falls quickly with radius,
stars are more likely to meet a clump which has a lower orbit, which causes the mean angular momentum change to be negative.
In a region where the clump density falls slowly,
the mean angular momentum change can be positive.
Simulations 
show that either type of the clump distribution can generate an exponential disc  thorough stellar scattering. Therefore, the origin of an exponential is independent of the sign of angular momentum changes.  \citet{Elmegreen2016ApJ...830..115E} showed how an exponential is reached when an inward scattering bias is present in a disc. Future theoretical work should be done on why
an exponential can emerge from scatterings with no bias or with an outward bias.



\section*{Data availability}

The data used in this article will be shared on request to the corresponding author.




\bibliographystyle{mnras}
\bibliography{reference} 








\bsp	
\label{lastpage}
\end{document}